\documentstyle[preprint,aps,tighten,epsfig]{revtex}
\begin{document}
\draft
\renewcommand{\thefootnote}{\fnsymbol{footnote}}
\title{Coherent and incoherent chaotic tunneling near singlet-doublet
crossings}
\author{Sigmund Kohler, Ralf Utermann, and Peter H\"anggi}
\address{Institut f{\"ur} Physik, Universit{\"a}t Augsburg,
      Memminger Stra{\ss}e 6, D--86135 Augsburg, Germany}
\author{Thomas Dittrich\footnote{Present address: Division de Physique
Th{\'e}orique, Institut de Physique Nucl{\'e}aire, F--91406 Orsay
Cedex, France}}
\address{Max-Planck-Institut f\"ur Physik komplexer Systeme, N{\"o}thnitzer
      Stra{\ss}e 38, D--01187 Dresden, Germany}
\date{\today}
\maketitle
%
\begin{abstract}
In the spectrum of systems showing chaos-assisted tunneling, three-state
crossings are formed when a chaotic singlet intersects a tunnel doublet. We
study the dissipative quantum dynamics in the vicinity of such crossings. A
harmonically driven double well coupled to a bath serves as a model. Markov
and rotating-wave approximations are introduced with respect to the Floquet
spectrum of the time-dependent central system. The resulting master equation
is integrated numerically. We find various types of transient tunneling,
determined by the relation of the level width to the inherent energy scales of
the crossing. The decay of coherent tunneling can be significantly retarded or
accelerated. Modifications of the quantum asymptotic state by the crossing are
also studied. The comparison with a simple three-state model shows that in
contrast to the undamped case, the participation of states outside the
crossing cannot be neglected in the presence of dissipation.
\end{abstract}
\section{Introduction}
\label{sec:intro}
\begin{figure}
\centerline{
\psfig{width=8cm,figure=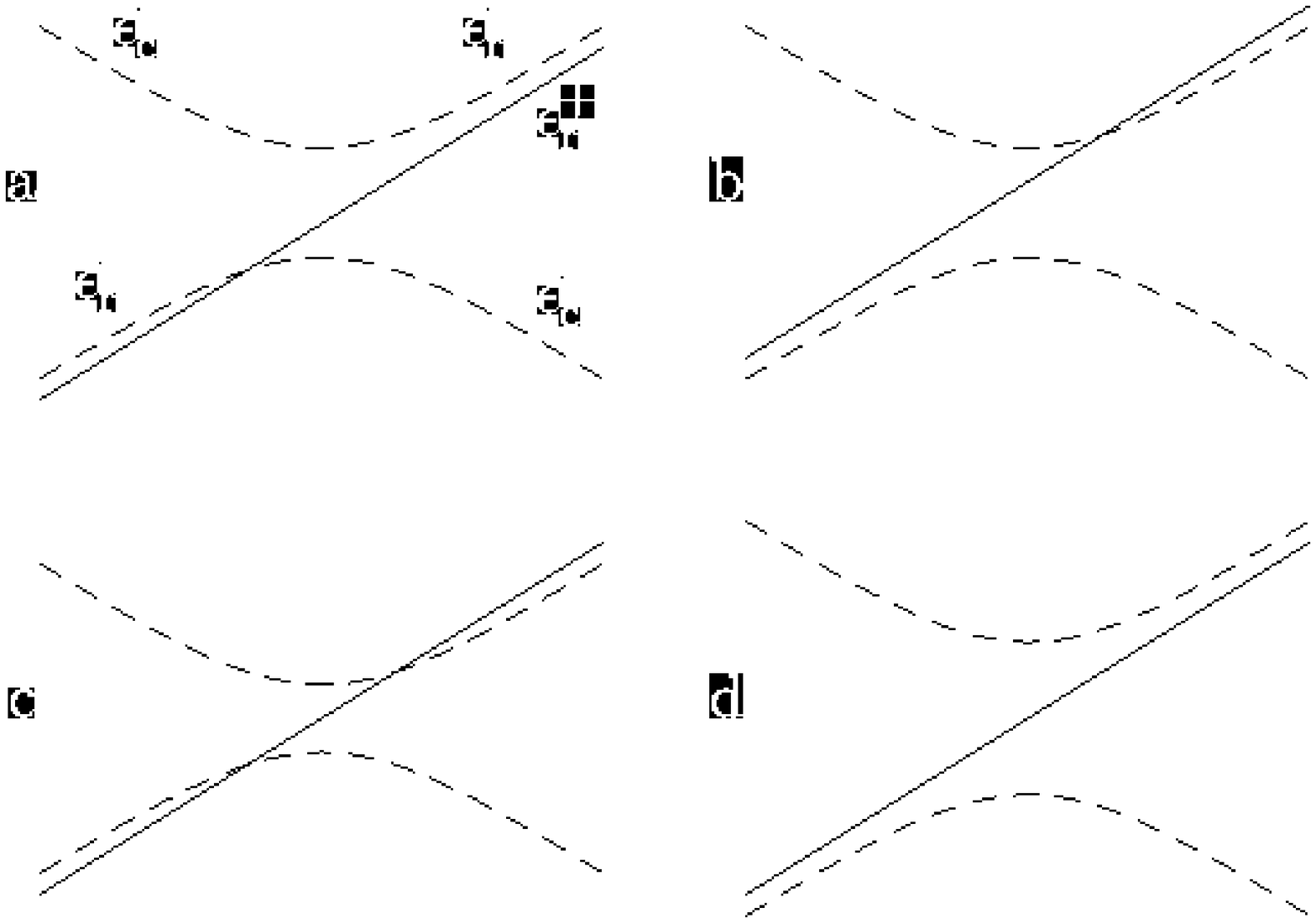}
}
\caption{
Possible configurations of quasienergy crossings between a chaotic singlet and
a regular doublet. Different line types signify different parity. See
Sect.~\protect\ref{sec:3s} for the labeling of the levels. Note that only for
configurations a,b, the order of the regular doublet is restored in passing
through the crossing. In configurations c,d, it is reversed.
}
\label{fig:crossing}
\end{figure}

In a quantum setting, the coexistence of regular and chaotic regions in a
mixed phase space leads to a variety of uncommon coherence phenomena. Most
prominent among them is chaotic
tunneling\cite{dav,boh1,lin,pla,boh2,tom,ute1,ute2,ron,lat,dor,fri,zan,ley},
the coherent exchange of probability between symmetry-related regular islands
that are separated by a chaotic layer, not by a static potential
barrier. Chaotic tunneling comes about by an interplay of classical
nonlinear---typically bistable---dynamics and quantum coherence. It therefore
reflects features of the classical phase space, such as the width of the
chaotic layer\cite{ute1,ute2} and the structure of the ``coast line''
\cite{dor,fri} separating it from the adjacent regular regions, as well as
fine details of the quantum spectrum like exact and avoided crossings. In
previous work \cite{ute1,ute2}, it has been shown that, as the chaotic layer
grows with increasing nonlinearity, the tunnel splittings widen collectively.
Superimposed on this global trend of the parameter
dependence, however, there are strong fluctuations, occurring on a smaller
parameter scale and restricted to individual tunnel doublets
\cite{tom,dor,zan,ley}.  A major source of these
fluctuations are the disturbances of doublets, suffered as they are
intersected by other levels. The most common type of such intersections is
formed when a doublet, pertaining to a pair of eigenstates located on
symmetry-related quantizing tori, encounters a singlet that belongs to an
eigenstate in the chaotic sea\cite{lat}.
The chaotic states, even if they no longer come in pairs close in energy,
can still be classified as even or odd. This fact determines the structure
of the singlet-doublet crossings: The
partner in the doublet sharing the symmetry of the chaotic singlet is repelled
by the singlet---together, they form an avoided crossing. The partner with
opposite symmetry must either be intersected in an exact crossing close to the
avoided one, or else the order within the doublet is reversed from one side of
the crossing to the other. The variety of configurations of such crossings is
sketched in Fig.~\ref{fig:crossing}.

In the present paper, we study chaotic tunneling in the vicinity of
singlet-doublet crossings, under the influence of incoherent processes. Near a
crossing, level separations deviate vastly, in both directions, from the
typical tunnel splitting (cf.\ Fig.~\ref{fig:crossing}). This is reflected in
time-domain phenomena ranging from the suppression of tunneling to a strong
increase in its rate and to complicated quantum beats\cite{lat}. In Section
\ref{sec:cons}, we briefly review chaotic tunneling in our working model, a
harmonically driven quartic double well\cite{gro1,gro2,gro3}. Singlet-doublet
crossings are identified and characterized by their signature in terms of
quasienergy and mean energy. We describe the behaviour of the eigenstates
close to a crossing and analyze the coherent dynamics in terms of a simple
three-state model.

Tunneling is associated with extremely small energy scales, all the more in
the semiclassical regime we are interested in. It is therefore particularly
sensitive to any disruption of coherence as it occurs due to the unavoidable
coupling to the environment. As immediate consequences, the symmetry
underlying the formation of tunnel doublets is generally broken, and an
additional energy scale is introduced, the effective finite width attained by
each discrete level. Tunneling and related coherence phenomena are thus
rendered transients that occur---if at all---on the way towards an asymptotic
equilibrium state. Singlet-doublet crossings, in turn, drastically change the
nondissipative energy scales and replace the two-level by a three-level
structure. As a consequence, the familiar way tunneling fades out in the
presence of dissipation is also significantly altered. Near a crossing, the
coherent dynamics can last much longer than for the unperturbed doublet,
despite the presence of the same decoherence than outside the crossing,
establishing ``chaos-induced coherence''. Depending on temperature, it can
also be destroyed on a much shorter time scale.

In Section \ref{sec:diss}, we first outline the modifications of our model
necessary to incorporate dissipation. We treat the quantum dissipative
dynamics within the framework of the Floquet formalism
\cite{dit1,dit2,blu1,blu2,gra,koh}. In the present context of an interplay of
tunneling with incoherent processes, we focus on the regime of weak
dissipation. Very long time scales are therefore involved, both in the
coherent dynamics and in the bath response. In this case, the usual (Markov
and rotating-wave) approximations resorted to in the elimination of the
external degrees of freedom, have to be reconsidered critically.  The coupling
to the environment indirectly couples the three levels in the crossing to all
the other states of the central system. On the basis of numerical results for
the full driven double well with dissipation, we reveal the limitations of the
three-level approximation and identify additional features of the full
dynamics not covered by it. In particular, we consider the long-time
asymptotics---the quantum attractor---and the phase-space structure associated
with it. Section \ref{sec:conc} serves to summarize our results and to suggest
directions of further research.
\section{Conservative classical and quantum dynamics}
\label{sec:cons}
\subsection{The model and its symmetries}
\label{sec:symm}
As a prototypical working model, we consider the quartic double well with a
spatially homogeneous driving force harmonic in time. It is defined by the
Hamiltonian
\begin{eqnarray}
\label{eq:doublewell}
H_{\rm DW}(x,p;t) &=& H_0(x,p) + H_1(x;t),\nonumber\\
H_0(x,p) &=& {p^2 \over 2} - {1 \over 4} x^2 +{1 \over 64 D} x^4, \\
H_1(x;t) &=& S\,x\,{\rm cos}(\omega t).\nonumber
\end{eqnarray}
Apart from mere scaling, the classical phase space of $H_0(x,p)$ depends only
on the presence or absence, and the sign, of the $x^2$ term. Otherwise, it has
no free parameter. In the quantum-mechanical case, the parameter $D/\hbar$
plays the r{\^o}le of an inverse quantum of action. It controls the barrier
height and can be interpreted as the (approximate) number of doublets with
energies below the top of the barrier. Accordingly, the classical limit is
reached by letting $D/\hbar\rightarrow\infty$. The influence of the driving on
the classical phase space structure is characterized by the rescaled amplitude
\begin{equation}
F=S/\sqrt{8D}
\end{equation}
and frequency $\omega$. This implies that the classical dynamics
is independent of the barrier height $D$.

The unperturbed Hamiltonian $H_0(x,p)$ is invariant under the parity ${\sf
P}$: $x \rightarrow -x$, $p \rightarrow -p$, $t \rightarrow t$. This symmetry
is generally destroyed by the driving. With the above choice of $H_1(x;t)$,
however, a more general, dynamical symmetry is retained
\cite{gro1,gro2,gro3,per}. It is defined by the operation
\begin{equation}
{\sf P_{\omega}}:\quad  x \rightarrow -x, \quad p \rightarrow -p, \quad
t \rightarrow t + \pi/\omega,
\label{parity}
\end{equation}
and represents a generalized parity acting in the extended phase space spanned
by $x$, $p$, and phase, i.e., time $t\,{\rm mod}\, (2\pi / \omega)$. The
invariance under ${\sf P_{\omega}}$ of $H(x,p;t)$ allows to classify all its
quantum eigenstates as even or odd.

As a consequence of the periodic time dependence of $H(x,p;t)$, the relevant
generator of the quantum dynamics is now the Floquet operator
\cite{shi,sam,man,chu,hae1}
\begin{equation}
U = {\cal T}\, {\rm exp} \left( -{{\rm i} \over \hbar}
\int_0^{2\pi / \omega} {\rm d} t \, H(t) \right),
\end{equation}
where ${\cal T}$ denotes time ordering. According to the Floquet theorem, the
adiabatic states of the system are the eigenstates of $U$. They can be written
in the form
\begin{equation}
|\psi_{\alpha}(t)\rangle = {\rm e}^{-{\rm i} \epsilon_{\alpha}t/\hbar}
|\phi_{\alpha}(t)\rangle, \quad
\label{eq:floquetstat}
\end{equation}
with
\begin{eqnarray*}
|\phi_{\alpha}(t + 2\pi / \omega)\rangle = |\phi_{\alpha}(t)\rangle.
\end{eqnarray*}
Expanded in the basis spanned by these Floquet states, the propagator of the
driven system reads
\begin{equation}
U(t',t)=\sum_\alpha {\rm e}^{-{\rm i}\epsilon_\alpha (t'-t)/\hbar}
|\phi_\alpha(t')\rangle\langle\phi_\alpha(t)|.
\label{eq:floquetprop}
\end{equation}
The associated eigenphases $\epsilon_{\alpha}$, referred to as quasienergies,
come in classes, $\epsilon_{\alpha,n}=\epsilon_{\alpha}+n\hbar\omega$, $n = 0,
\pm 1, \pm 2, \ldots$. This is suggested by a Fourier expansion of the
$|\phi_{\alpha}(t)\rangle$,
\begin{eqnarray}
{\displaystyle |\phi_{\alpha}(t)\rangle} &=&
{\displaystyle \sum_n |c_{\alpha,n}\rangle\,
{\rm e}^{-{\rm i} n\omega t},} \nonumber \\
{\displaystyle |c_{\alpha,n}\rangle} &=&
{\displaystyle {\omega \over 2\pi} \int_0^{2\pi / \omega} {\rm d} t \,
|\phi_{\alpha}(t)\rangle\,{\rm e}^{{\rm i} n\omega t}.}
\label{floquet:fourier}
\end{eqnarray}
The index $n$ counts the number of quanta in the driving field. Otherwise the
members of a class $\alpha$ are physically equivalent. Therefore, the
quasienergy spectrum can be reduced to a single ``Brillouin zone'',
$-\hbar\omega/2 \leq \epsilon < \hbar\omega/2$.

Since the quasienergies have the character of phases, they can be ordered only
locally, not globally. A quantity that is defined on the full real axis and
therefore does allow for a complete ordering, is the mean energy
\cite{chu,hae1}
\begin{equation}
E_{\alpha}
= {\omega \over 2\pi} \int_0^{2\pi / \omega} {\rm d} t \,
\langle\psi_{\alpha}(t)|\,H(t)\,|\psi_{\alpha}(t)\rangle
\equiv \langle\langle\phi_{\alpha}(t)|\,H(t)\,|
\phi_{\alpha}(t)\rangle\rangle.
\label{eq:meanen}
\end{equation}
It is related to the corresponding quasienergy by
\begin{equation}
E_{\alpha} = \epsilon_{\alpha} + \langle\langle\phi_{\alpha}(t)
|\,{\rm i}\hbar\frac{\partial}{\partial t}\,|\phi_{\alpha}(t)\rangle\rangle,
\end{equation}
where the outer angle brackets denote the time average over one period of the
driving, as indicated in eq.~(\ref{eq:meanen}). The second term on the
right-hand side plays the r{\^o}le of a geometric phase accumulated over this
period\cite{moo}. Without the driving, $E_{\alpha} = \epsilon_{\alpha}$, as
it should be. By inserting the Fourier expansion (\ref{floquet:fourier}) the
mean energy takes the form
\begin{equation}
E_\alpha = \sum_n(\epsilon_\alpha+n\hbar\omega)\,
\langle c_{\alpha,n}|c_{\alpha,n}\rangle.
\label{eq:meanen:fourier}
\end{equation}
It shows that the $n\,$th Floquet channel gives a contribution
$\epsilon_\alpha + n\hbar\omega$ to the mean energy, weighted by the Fourier
coefficient $\langle c_{\alpha,n}|c_{\alpha,n}\rangle$\cite{hae1}.

Quasienergies and Floquet states are obtained numerically by solving the
matrix eigenvalue equation\cite{shi,chu,hae1}
\begin{equation}
\sum_{n'} \sum_{k'} H_{n,k;n',k'} c_{n',k'} = \epsilon c_{n,k},
\label{eq:floqueteq}
\end{equation}
equivalent to the time-dependent Schr\"odinger equation. It is derived by
inserting the eigenstates (\ref{eq:floquetstat}) into the Schr\"odinger
equation, Fourier expanding, and using the representation by the eigenstates
of the unperturbed Hamiltonian, $H_0 |\Psi_k\rangle = E_k
|\Psi_k\rangle$. We introduced the abbreviations
\begin{eqnarray*}
H_{n,k;n',k'} &=& (E_k - n\hbar\omega) \delta_{n-n'} \delta_{k-k'} \\
&&+\frac{1}{2}\,S\, x_{k,k'} \, (\delta_{n-1-n'} + \delta_{n+1-n'}), \\
c_{n,k} &=& \langle\Psi_k|c_n\rangle, \\
x_{k,k'} &=& \langle\Psi_{k}|\,x\,|\Psi_{k'}\rangle. 
\end{eqnarray*}
The invariance of the system under ${\sf P_{\omega}}$ is of considerable help
in solving Eq.~(\ref{eq:floqueteq}), because it completely decouples the
respective systems of eigenvalue equations for the even and odd subspaces
\cite{ute1,ute2}.

\subsection{Coherent chaotic tunneling}
\label{sec:chao}
With the driving $H_1(x;t)$ switched off, the classical phase space generated
by $H(x,p;t)$ exhibits the constituent features of a bistable Hamiltonian
system. There is a separatrix at $E = 0$. It forms the border between two sets
of trajectories: One set, with $E < 0$, comes in symmetry-related pairs, each
partner of which oscillates in either one of the two potential minima. The
other set consists of unpaired trajectories, with $E > 0$, that encircle both
wells in a spatially symmetric fashion.

\begin{figure}
\centerline{
\psfig{width=8cm,figure=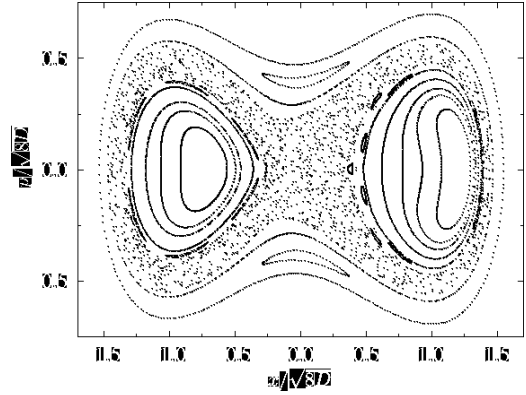}
}
\vspace{0.5cm}
\caption{
Stroboscopic classical phase-space portraits, at $\omega t=2\pi n$, of the
harmonically driven quartic double well, Eq.~(\protect\ref{eq:doublewell}),
at $F=0.015$, $\omega=0.982$.
}
\label{fig:cldyn}
\end{figure}

Increasing the amplitude of the driving from zero onwards has two principal
consequences for the classical dynamics (Fig.~\ref{fig:cldyn}): The separatrix
is destroyed as a closed curve and replaced by a homoclinic tangle\cite{lic}
of stable and unstable manifolds. As a whole, it forms a chaotic layer in the
vicinity and with the topology of the former separatrix. It opens the way for
diffusive transport between the two potential wells. Due to the nonlinearity
of the potential, there is an infinite set of resonances of the driving with
the unperturbed motion, both inside and outside the wells\cite{esc,rei1}.
Since the period of the unperturbed, closed trajectories diverges for $E \to
0$, the resonances accumulate towards the separatrix of the unperturbed
system. By its sheer phase-space area, the first resonance (the one for which
the periods of the driving and of the unperturbed oscillation are in a ratio
of 1:1) is prominent among the others and soon (in terms of increasing
amplitude $F$) exceeds the size of the ``order-zero'' regular areas near the
bottom of each well. For the values of $F$ and $\omega$ chosen in the
numerical calculations in this paper, all higher resonances remain negligible
in size. The borderline between the chaotic layer along the former separatrix
and the regular regions within and outside the wells is therefore quite
sharply defined. The ``coastal strip'' formed by hierarchies of regular
islands around higher resonances remains narrow. For the tunneling dynamics,
the r\^ole of states located in the border region\cite{dor} is therefore not
significant here.

\begin{figure}
\centerline{
\psfig{width=8cm,figure=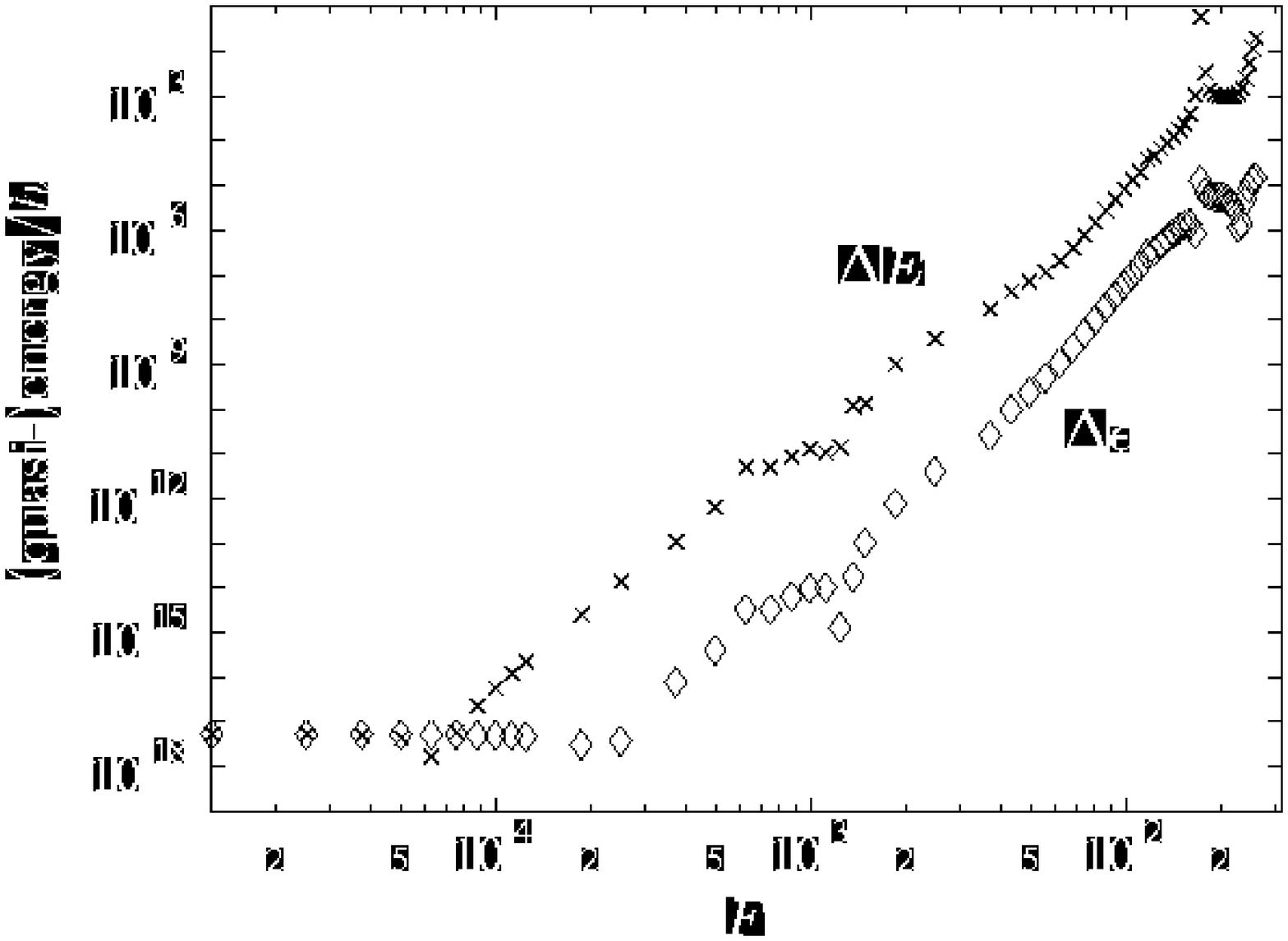}
}
\caption{
Dependence on the driving amplitude of the splitting of the ground-state
doublet, of the harmonically driven quartic double well,
Eq.~(\protect\ref{eq:doublewell}), at $D/\hbar=8$ and $\omega=0.95$. Diamonds
are quasienergy, crosses mean-energy splittings. The dips interrupting the
smooth parameter dependence indicate crossings of the ground-state doublet
with chaotic singlets. After \protect\cite{ute2}.
}
\label{fig:powerlaw}
\end{figure}

Both major tendencies in the evolution of the classical phase
space---extension of the chaotic layer and growth of the first
resonance---leave their specific traces in the quasienergy spectrum. The
tunnel doublets characterizing the unperturbed spectrum for $E < 0$ pertain to
states located on pairs of symmetry-related quantizing tori in the regular
regions within the wells. With increasing size of the chaotic layer, the
quantizing tori successively resolve in the chaotic sea. The corresponding
doublets disappear as distinct structures in the spectrum as they attain a
splitting of the same order as the mean level separation. The gradual widening
of the doublets proceeds as a smooth function of the driving amplitude
\cite{ute1,ute2}. This function roughly obeys a power law\cite{wil}, cf.\
Fig.~\ref{fig:powerlaw}. As soon as a pair of states is no longer supported by
any torus-like manifold, including fractal\cite{rei2} and vague tori
\cite{rein}, the corresponding eigenvalues detach themselves from the regular
ladder to which they formerly belonged. They can then fluctuate freely in the
spectrum and thereby ``collide'' with other chaotic singlets or regular
doublets.

The appearance of a regular region, large enough to accomodate several
eigenstates, around the first resonance introduces a second ladder of doublets
into the spectrum. Size and shape of the first resonance vary in a way
different from the fate of the main regular region. The corresponding doublet
ladder therefore moves in the spectrum independently of the doublets that
pertain to the main regular region, and of the chaotic singlets. This gives
rise to additional singlet-doublet and even to doublet-doublet encounters.

\subsection{Three-level crossings}
\label{sec:3s}
Among the various types of quasienergy crossings that occur according to the
above scenario, those involving a regular doublet and a chaotic singlet are
the most common. In order to give a quantitative account of such crossings and
the associated coherent dynamics, and for later reference in the context of
the incoherent dynamics, we shall now discuss them in terms of a simple
three-state model, devised much in the spirit of Ref.~\cite{boh2}.

Far to the left of the crossing, we expect the following situation: There is a
doublet of Floquet states
\begin{eqnarray}
|\psi_{\rm r}^+(t)\rangle
&=& {\rm e}^{-{\rm i}\epsilon_{\rm r}^+ t/\hbar}|\phi_{\rm r}^+(t)\rangle , \\
|\psi_{\rm r}^-(t)\rangle
&=& {\rm e}^{-{\rm i}(\epsilon_{\rm r}^+ +\Delta) t/\hbar}
|\phi_{\rm r}^-(t)\rangle ,
\end{eqnarray}
with even (superscript `$+$') and odd (`$-$') generalized parity,
respectively, residing on a pair of quantizing tori in one of the regular
(subscript `r') regions. We have assumed that the quasienergy splitting (as
opposed to the unperturbed splitting) is $\epsilon_{\rm r}^- - \epsilon_{\rm
r}^+ = \Delta > 0$. The global relative phases can be chosen such that
the superpositions
\begin{equation}
\label{eq:rightleft}
|\psi_{\rm R,L}(t)\rangle = \frac{1}{\sqrt{2}}\left(|\psi_{\rm r}^+(t)\rangle
\pm |\psi_{\rm r}^-(t)\rangle\right)
\end{equation}
are initially localized in the right and the left well, respectively, and
tunnel back and forth with a frequency $\Delta/\hbar$ given by the tunnel
splitting in the presence of the driving.

As the third player, we introduce a Floquet state
\begin{equation}
|\psi_{\rm c}^-(t)\rangle
= {\rm e}^{-{\rm i}(\epsilon_{\rm r}^+ + \Delta + \Delta_{\rm c})t/\hbar}
|\phi_{\rm c}^-(t)\rangle,
\end{equation}
located mainly in the chaotic (subscript `c') layer, so that its time-periodic
part $|\phi_{\rm c}^-(t)\rangle$ contains a large number of harmonics.
Without loss of generality, its generalized parity is fixed to be odd. For the
quasienergy, we have assumed that $\epsilon_{\rm c}^- = \epsilon_{\rm r}^+ +
\Delta + \Delta_{\rm c}$, where $|\Delta_{\rm c}|$ can be regarded as a
measure of the distance from the crossing.

The structure of the classical phase space then implies that the mean energy
of the chaotic state should be close to the top of the barrier and far above
that of the doublet. We assume, like for the quasienergies, a small splitting
of the mean energies pertaining to the regular doublet, $E_{\rm r}^- - E_{\rm
r}^+ \ll E_{\rm c}^- - E_{\rm r}^\pm$.

\begin{figure}
\centerline{
\psfig{width=8cm,figure=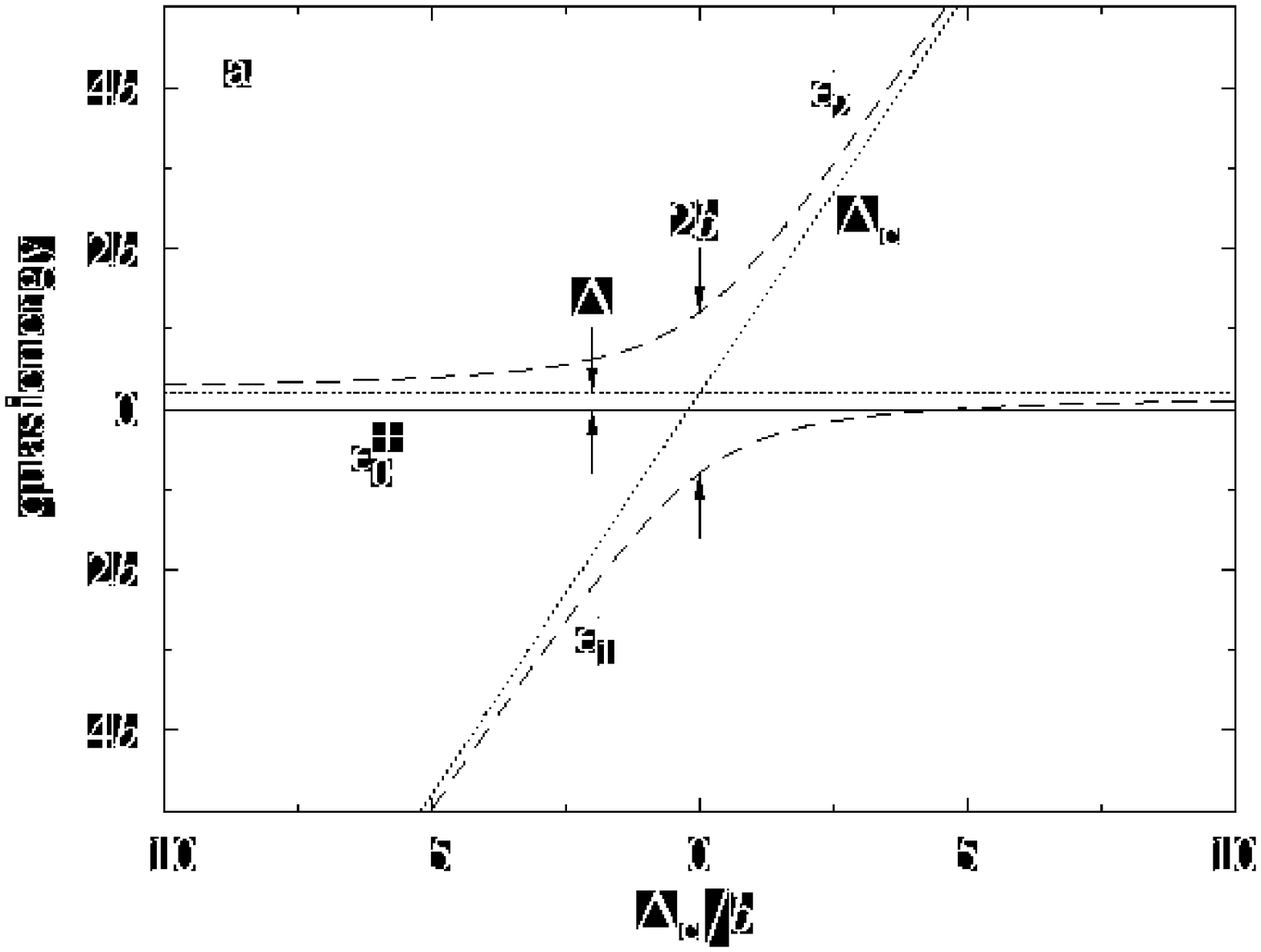}
\psfig{width=8cm,figure=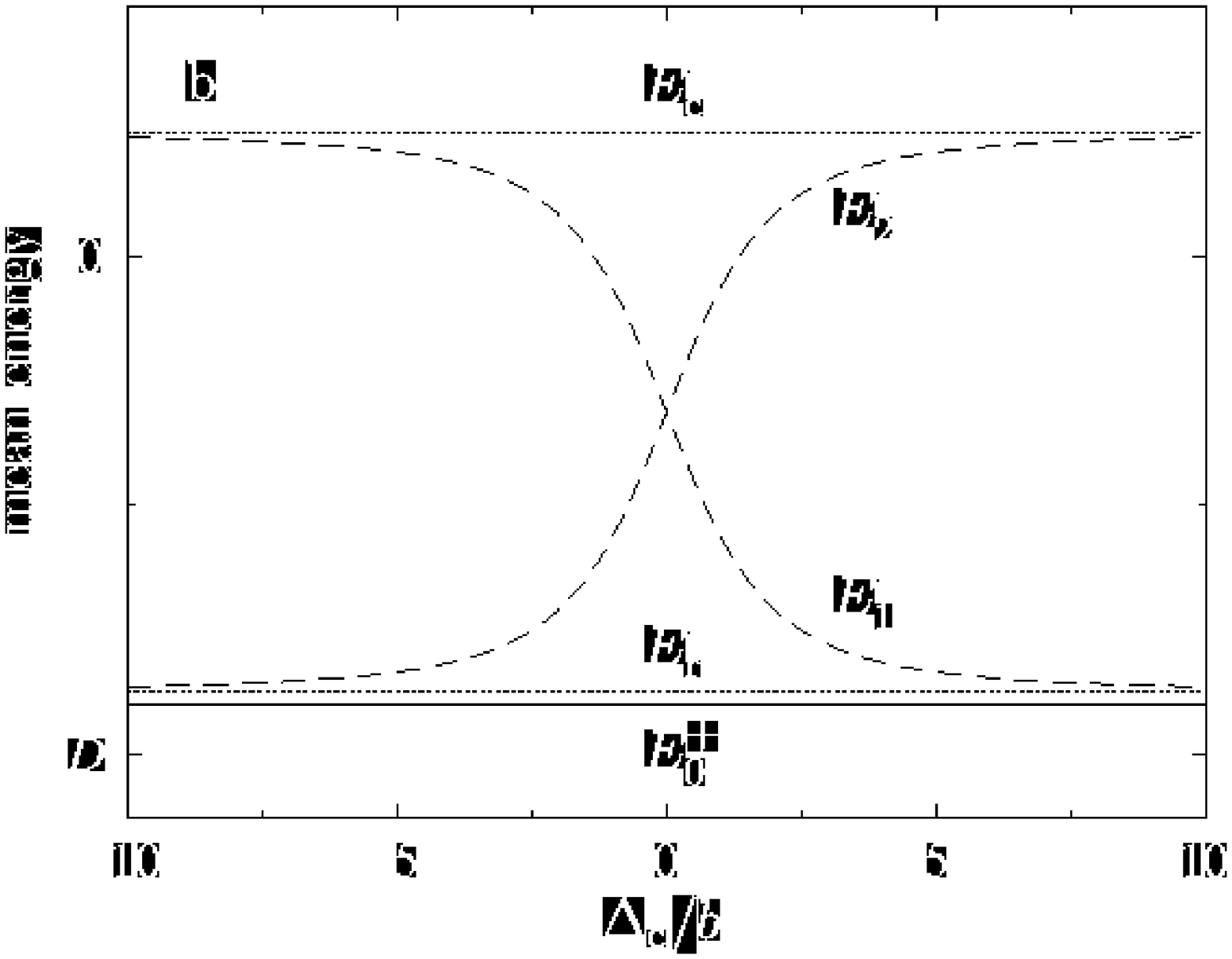}
}
\vspace{0.5cm}
\caption{
A singlet-doublet crossing, according to a three-state model,
Eq.~(\protect\ref{eq:ham3s}), in terms of the dependence of the quasienergies
(panel a) and the mean energies (b) on the coupling parameter
$\Delta_{\rm c}/b$. Unperturbed energies are marked by dotted lines, the
energies for the case with coupling by full lines for even and dashed lines
for odd states.
}
\label{fig:theo3s}
\end{figure}

In order to model an avoided crossing between $|\psi_{\rm r}^-\rangle$ and
$|\psi_{\rm c}^-\rangle$, we suppose that there is a non-vanishing matrix
element $\langle\langle\phi_{\rm r}^-|H_{\rm DW}|\phi_{\rm c}^-\rangle\rangle
= b > 0$. For the singlet-doublet crossings under study, we typically find
that $\Delta \ll b \ll \hbar\omega$. Neglecting the coupling with all
other states, we obtain the three-state (subscript `3s') Floquet Hamiltonian
\begin{equation}
{\cal H}_{\rm 3s} = \epsilon_{\rm r}^+
+\left(\begin{array}{ccc}
        0 & 0      &                     0 \\
	0 & \Delta &                     b \\
	0 & b      & \Delta+\Delta_{\rm c}
\end{array}\right) ,
\label{eq:ham3s}
\end{equation}
in the three-dimensional Hilbert space spanned by $\{|\phi_{\rm
r}^+(t)\rangle, |\phi_{\rm r}^-(t)\rangle, |\phi_{\rm c}^-(t)\rangle\}$. Its
Floquet states read
\begin{eqnarray}
\label{eq:psi012}
|\psi_0^+(t)\rangle
&=& {\rm e}^{-{\rm i}\epsilon_0^+t/\hbar}
|\phi_{\rm r}^+(t)\rangle , \nonumber\\
|\psi_1^-(t)\rangle
&=& {\rm e}^{-{\rm i}\epsilon_1^- t/\hbar}
    \left( |\phi_{\rm r}^-(t)\rangle\cos\beta 
    - |\phi_{\rm c}^-(t)\rangle\sin\beta \right) , \\
|\psi_2^-(t)\rangle
&=& {\rm e}^{-{\rm i}\epsilon_2^- t/\hbar}\left( |\phi_{\rm r}^-(t)
    \rangle\sin\beta + |\phi_{\rm c}^-(t)\rangle\cos\beta \right) .\nonumber
\end{eqnarray}
Their quasienergies are
\begin{equation}
\epsilon_0^+ = \epsilon_{\rm r}^+,\quad
\epsilon_{1,2}^- = \epsilon_{\rm r}^+ + \Delta +
\frac{1}{2}\Delta_{\rm c}\mp\frac{1}{2} \sqrt{\Delta_{\rm c}^2+4b^2}.
\end{equation}
The mean energies are approximately given by
\begin{eqnarray}
\label{eq:meanen3s}
E_0^+ &=& E_{\rm r}^+ , \nonumber\\
E_1^- &=& E_{\rm r}^-\cos^2\beta + E_{\rm c}^-\sin^2\beta , \\
E_2^- &=& E_{\rm r}^-\sin^2\beta + E_{\rm c}^-\cos^2\beta , \nonumber
\end{eqnarray}
where contributions of the matrix element $b$ have been neglected.
The angle $\beta$ describes the mixing between the Floquet states $|\psi_{\rm
r}^-\rangle$ and $|\psi_{\rm c}^-\rangle$ and is a measure of the
distance to the avoided crossing. By diagonalizing the Hamiltonian
(\ref{eq:ham3s}), we obtain
\begin{equation}
2\beta = \arctan\left(\frac{2b}{\Delta_{\rm c}}\right), \quad
0 < \beta < \frac{\pi}{2}.
\end{equation}
For $\beta \to \pi/2$, corresponding to $-\Delta_{\rm c} \gg b$, we
retain the situation far left of the crossing, as outlined above, with
$|\psi_1^- \rangle \approx |\psi_{\rm c}^-\rangle$, $|\psi_2^-\rangle \approx
|\psi_{\rm r}^-\rangle$. To the far right of the crossing, i.e., for $\beta
\to 0$ or $\Delta_{\rm c} \gg b$, the exact eigenstates
$|\psi_1^-\rangle$ and $|\psi_2^-\rangle$ have interchanged their identity
with respect to the phase-space structure\cite{lat}. Here, we have
$|\psi_1^-\rangle \approx |\psi_{\rm r}^-\rangle$ and $|\psi_2^-\rangle
\approx |\psi_{\rm c}^-\rangle$.  The mean energy is essentially determined by
the phase-space structure. Therefore, there is also an exchange of $E_1^-$ and
$E_2^-$ in an exact crossing, cf.\ Eq.~(\ref{eq:meanen3s}), while $E_0^+$
remains unaffected (Fig.~\ref{fig:theo3s}b). The quasienergies $\epsilon_0^+$
and $\epsilon_1^-$ must intersect close to the avoided crossing of
$\epsilon_1^-$ and $\epsilon_2^-$ (Fig.~\ref{fig:theo3s}a). Their crossing is
exact, since they pertain to states with opposite parity (cf.\
Fig.~\ref{fig:crossing}a,b).

\begin{figure}
\centerline{
\psfig{width=8cm,figure=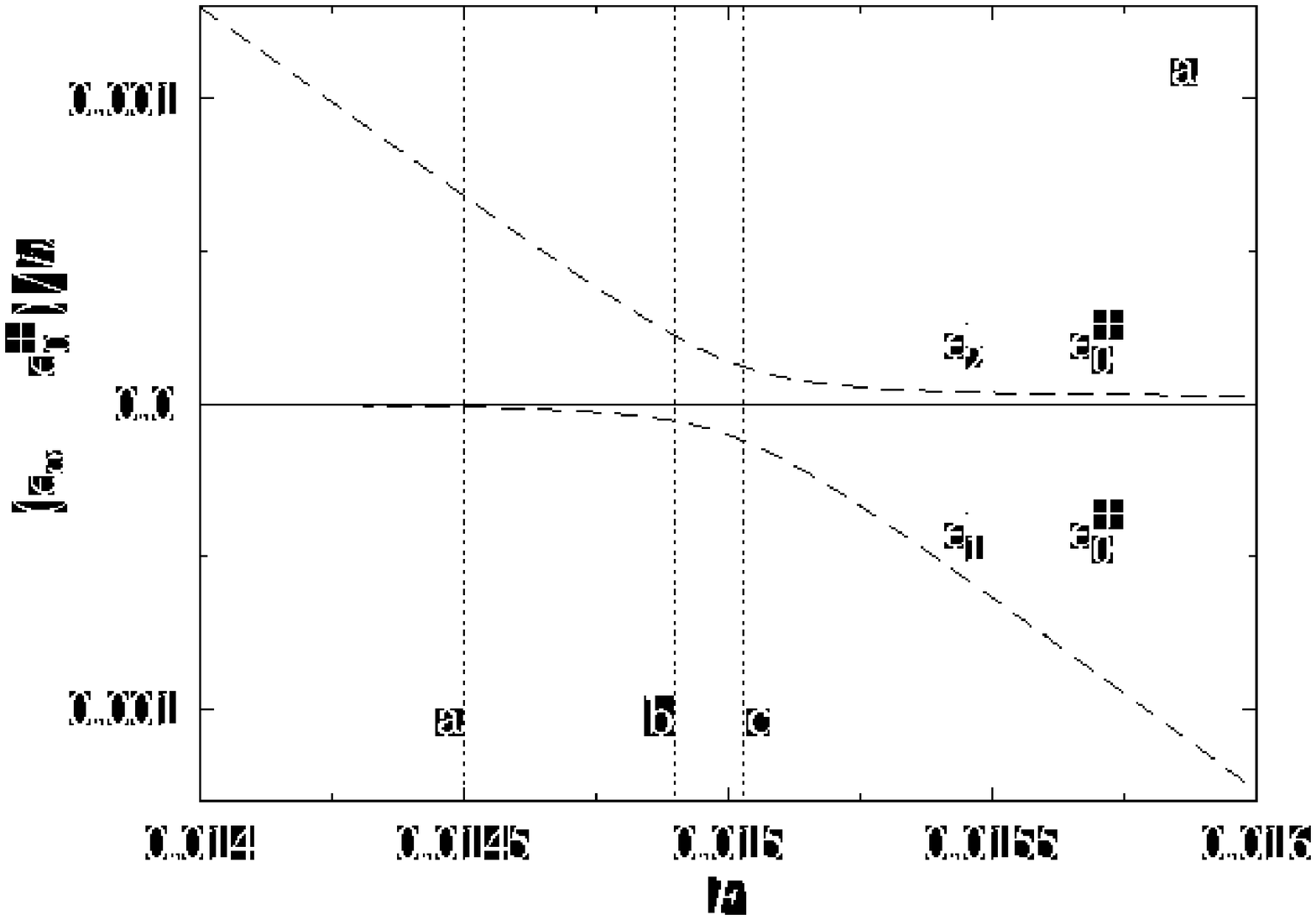}
\psfig{width=8cm,figure=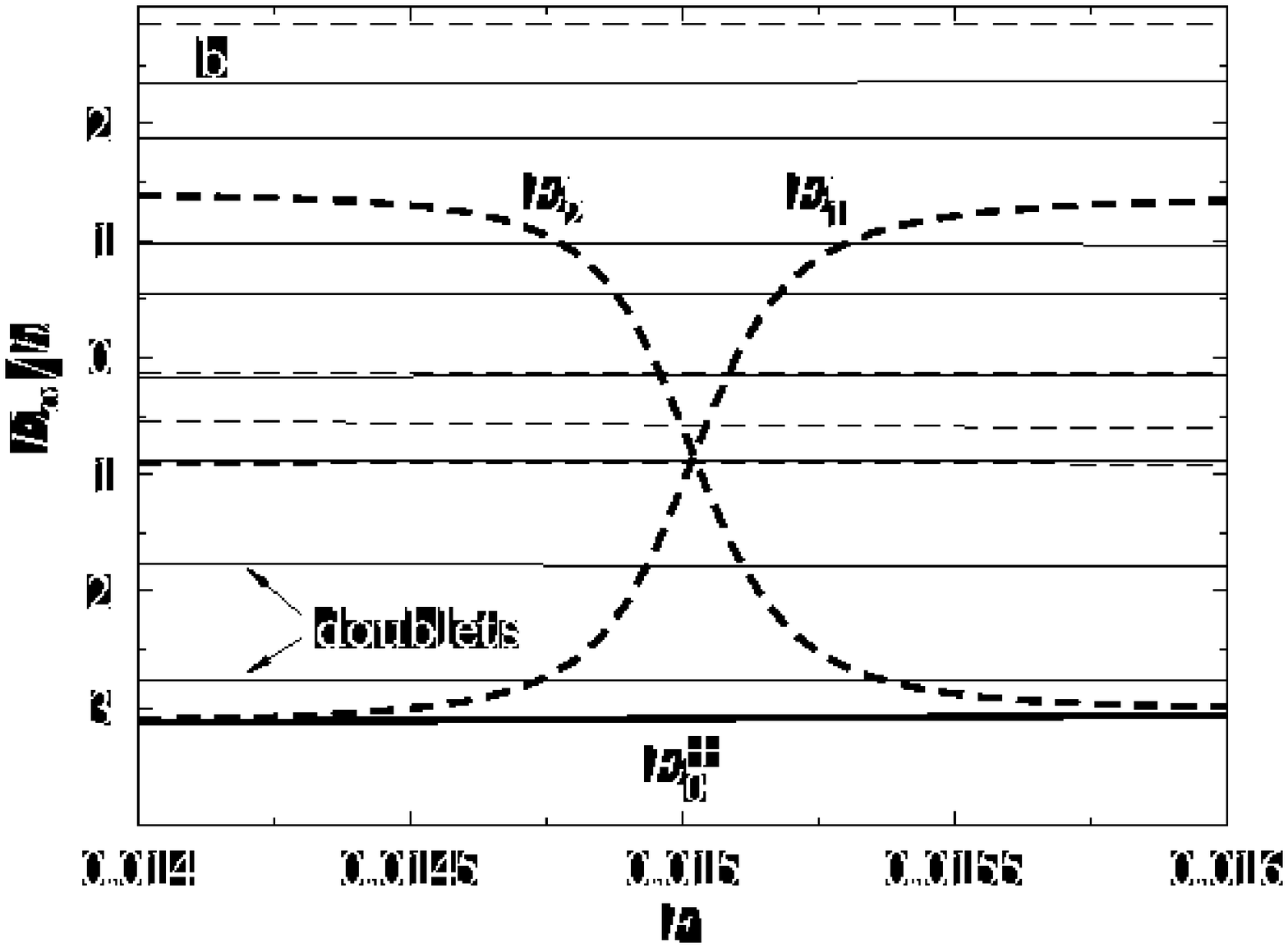}
}
\vspace{0.5cm}
\caption{
Singlet-doublet crossing found numerically for the driven double well,
Eq.~(\protect\ref{eq:doublewell}), at $D/\hbar = 4$ and $\omega = 0.982$, in
terms of the dependence of the quasienergies (a) and the mean energies (b) on
the driving amplitude $F = S/\protect\sqrt{8D}$. Values of the driving
amplitude used in Figs.~\protect\ref{fig:tun3s}
are marked by dotted vertical lines. Full and dashed lines indicate energies
of even and odd states, respectively. Bold lines give the mean energies of
the chaotic singlet and the ground-state doublet depicted in panel a.
}
\label{fig:num3s}
\end{figure}

\begin{figure}
\centerline{
\psfig{width=8cm,figure=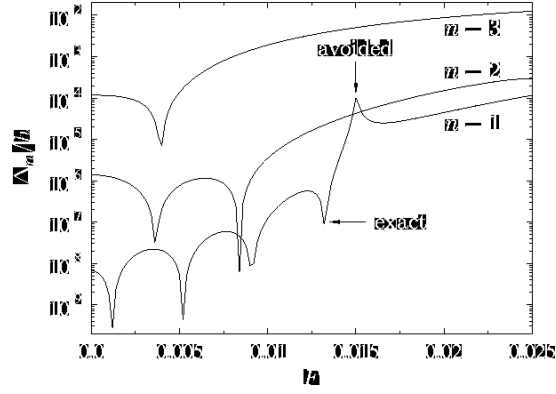}
}
\caption{
Splitting of the lowest doublets for $D/\hbar=4$ and $\omega=0.982$. The
arrows indicate the locations of the exact and the avoided crossing within a
three-level crossing of the type sketched in Fig.~\protect\ref{fig:crossing}a.
}
\label{fig:splitting}
\end{figure}

Numerical evidence shows that this idealized picture is not always correct. It
may well happen that even far away from a crossing, the doublet splitting does
not exactly return to its value on the opposite side
(cf.\ Figs.~\ref{fig:num3s}a, \ref{fig:splitting}).
It is even possible that an exact crossing of $\epsilon_0^+$ and
$\epsilon_1^-$ does not take place at all in the vicinity of the crossing. In
that case, the relation of the quasienergies in the doublet gets reversed via
the crossing (Fig.~\ref{fig:crossing}c,d). Nevertheless, the above scenario
captures the essential features.

To study the dynamics of the tunneling process, we focus on the state
\begin{equation}
|\psi(t)\rangle = \frac{1}{\sqrt{2}}\left(
{\rm e}^{-{\rm i}\epsilon_0^+ t/\hbar}|\phi_0^+(t)\rangle
+{\rm e}^{-{\rm i}\epsilon_1^- t/\hbar}|\phi_1^-(t)\rangle\cos\beta
+{\rm e}^{-{\rm i}\epsilon_2^- t/\hbar}|\phi_2^-(t)\rangle\sin\beta
\right).
\label{eq:tunnelstate}
\end{equation}
It is constructed such that at $t = 0$, it corresponds to the decomposition of
$(|\psi_{\rm r}^+\rangle + |\psi_{\rm r}^-\rangle)/\sqrt{2}$ (cf.\
Eq.~(\ref{eq:rightleft})) in the basis (\ref{eq:psi012}) at finite distance
from the crossing. Therefore, it is initially localized in the regular region
in the right well and follows the time evolution under the Hamiltonian
(\ref{eq:ham3s}). From Eqs.~(\ref{eq:rightleft}), (\ref{eq:psi012}),
we find the probabilities for its evolving into $|\psi_{\rm R}\rangle$,
$|\psi_{\rm L}\rangle$, or $|\psi_{\rm c}\rangle$, respectively, to be
\begin{eqnarray}
P_{\rm R}(t)
&=& |\langle\psi_{\rm R}(t)|\psi(t)\rangle|^2 \nonumber\\
&=& \frac{1}{2}\left(1 + \cos\frac{\epsilon_1t}{\hbar}\cos^2\beta+
    \cos\frac{\epsilon_2t}{\hbar}\sin^2\beta
    +\left[\cos\frac{(\epsilon_1-\epsilon_2)t}{\hbar}-1\right]
    \cos^2\beta\sin^2\beta\right), \nonumber\\
P_{\rm L}(t)
\label{eq:tun3s}
&=& |\langle\psi_{\rm L}(t)|\psi(t)\rangle|^2 \\
&=& \frac{1}{2}\left(1 - \cos\frac{\epsilon_1t}{\hbar}\cos^2\beta
    -\cos\frac{\epsilon_2t}{\hbar}\sin^2\beta
    +\left[\cos\frac{(\epsilon_1-\epsilon_2)t}{\hbar}-1\right]
    \cos^2\beta\sin^2\beta\right), \nonumber\\
P_{\rm c}(t) &=& |\langle\psi_{\rm c}(t)|\psi(t)\rangle|^2
=  \left[1-\cos\frac{(\epsilon_1-\epsilon_2)t}{\hbar}\right]
   \cos^2\beta\sin^2\beta. \nonumber
\end{eqnarray}

\begin{figure}
\centerline{
\psfig{width=8cm,figure=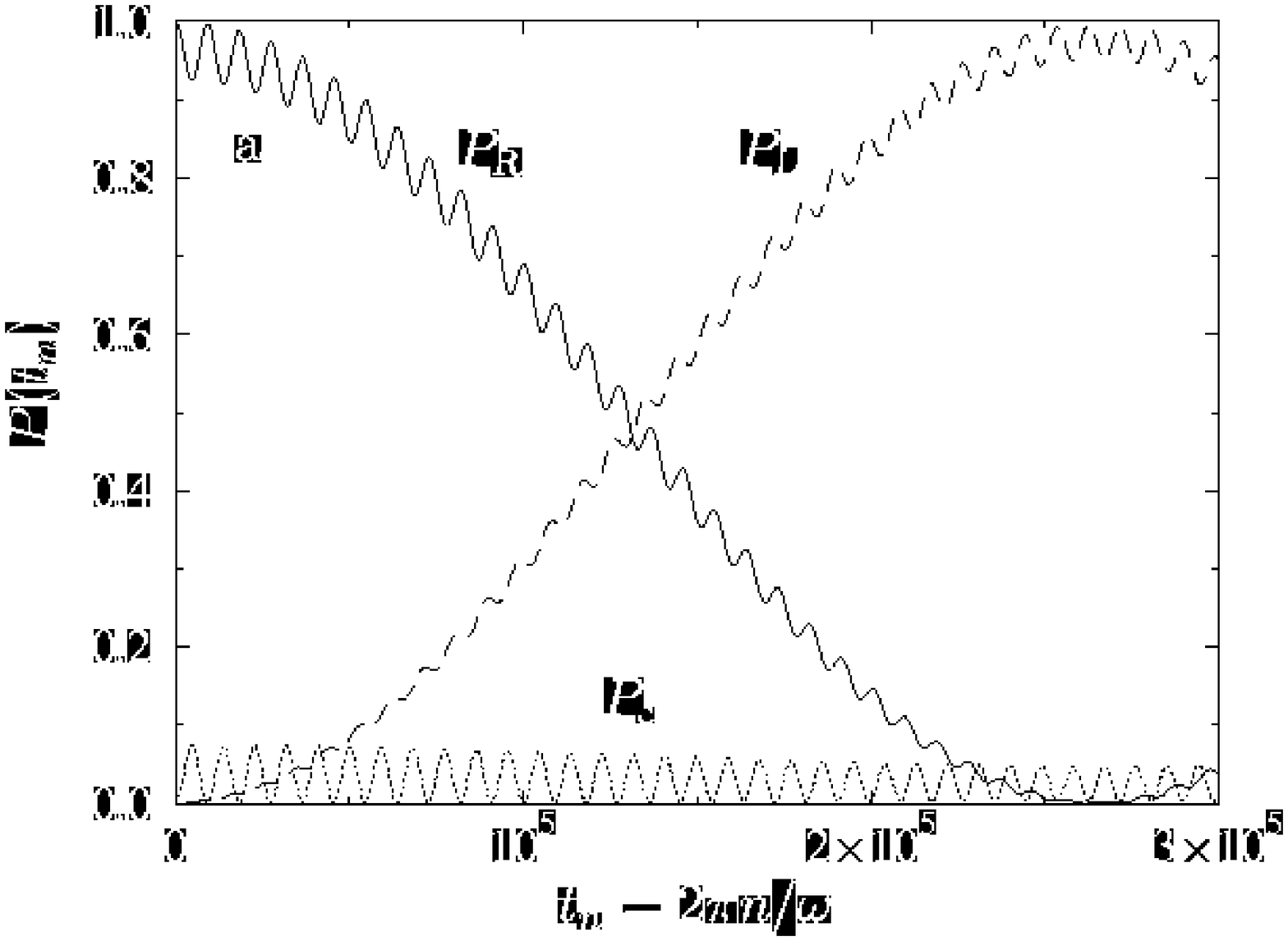}
\psfig{width=8cm,figure=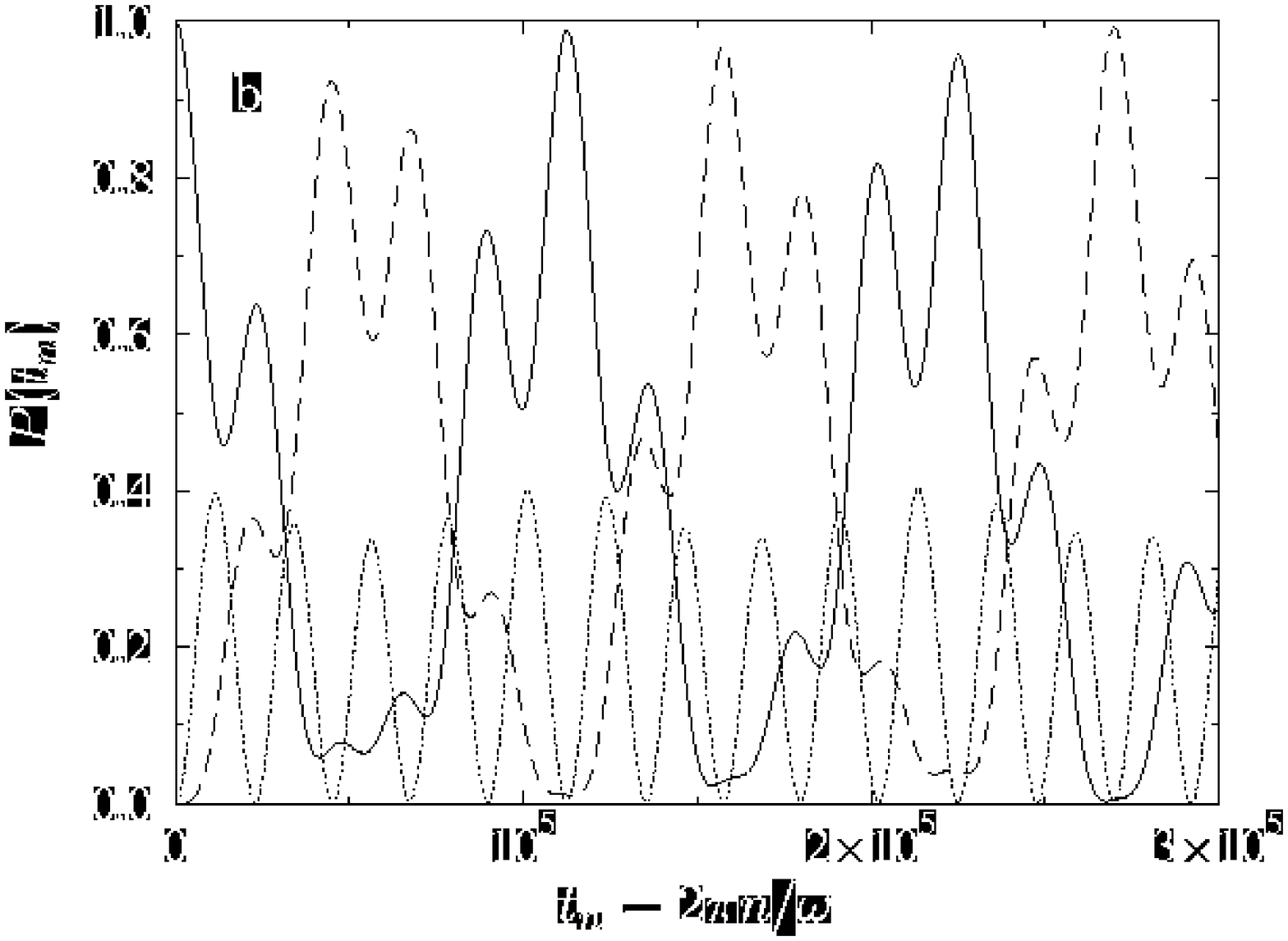}
}
\centerline{
\psfig{width=8cm,figure=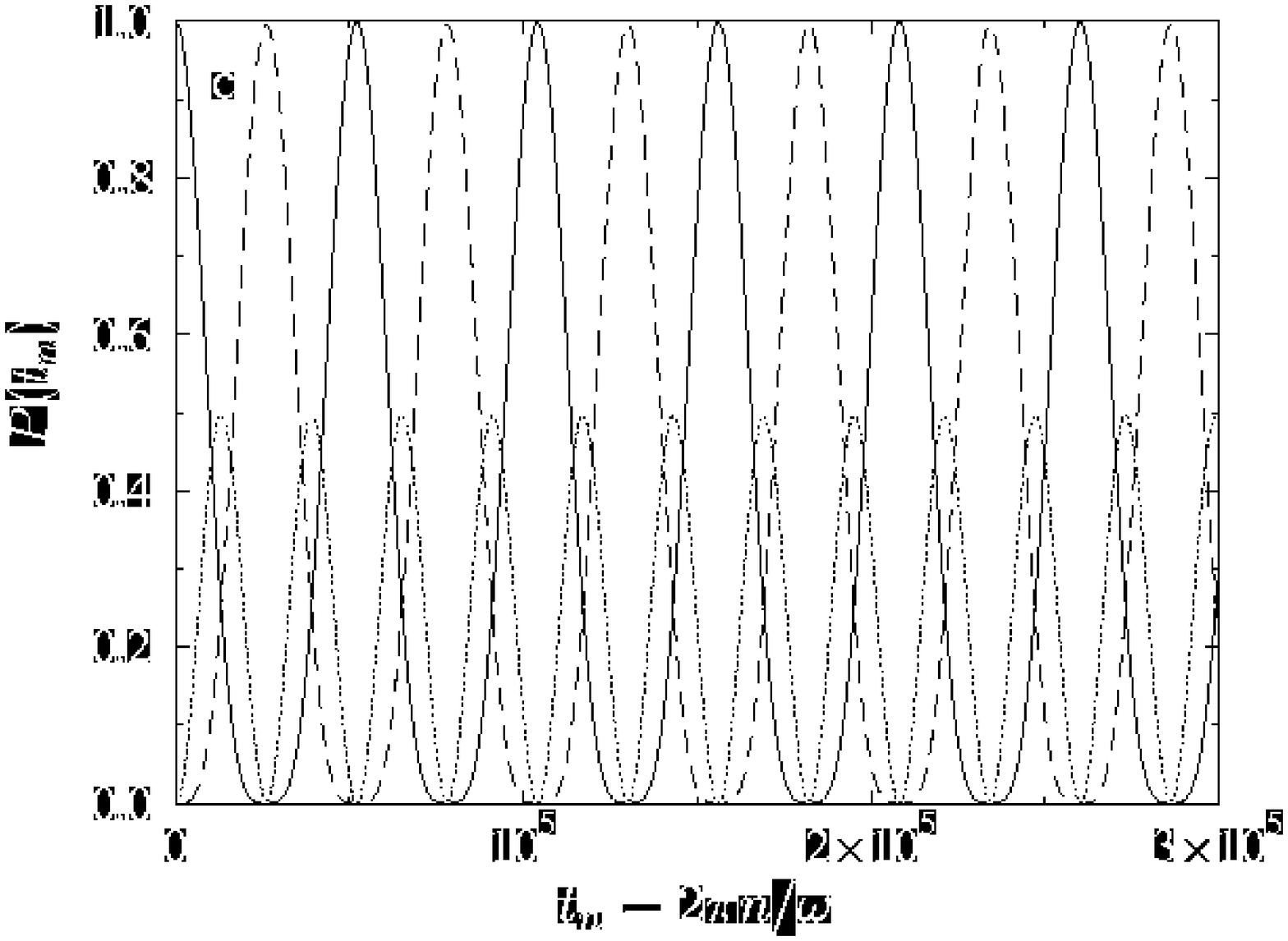}
}
\vspace{0.5cm}
\caption{
Time evolution of a state initially localized in the right well, in the
vicinity of the singlet-doublet crossing shown in
Fig.~\protect\ref{fig:num3s}, in terms of the probabilities to be in the right
well (return probability, marked by full lines), in the reflected state in the
left well (dashed), or in the chaotic state $|\psi_{\rm c}\rangle$ (dotted).
Parameter values are as in Fig.~\protect\ref{fig:num3s}, and $F = 0.0145$ (a),
$0.0149$ (b), $0.015029$ (c).
}
\label{fig:tun3s}
\end{figure}

In sufficient distance from the crossing, there is only little mixing between
the regular and the chaotic states, i.e., $\sin\beta\ll 1$ or $\cos\beta\ll
1$. The tunneling process then follows the familiar two-state dynamics
involving only $|\psi_{\rm r}^+\rangle$ and $|\psi_{\rm r}^-\rangle$, with
tunnel frequency $\Delta/\hbar$ (Fig.~\ref{fig:tun3s}a).

Close to the avoided crossing, $\cos\beta$ and $\sin\beta$ are of the same
order of magnitude, and $|\psi_1^-\rangle$, $|\psi_2^-\rangle$ become very
similar to one another. Both have now support in the chaotic layer as well as
in the symmetry-related regular regions and thus are of a hybrid nature. Here,
the tunneling involves all the three states and must at least be described by
a three-level system. The exchange of probability between the two regular
regions proceeds via a ``stop-over'' in the chaotic region\cite{boh2,lat}.
The three quasienergy differences that determine the time scales of this
process are in general all different, leading to complicated beats
(Fig.~\ref{fig:tun3s}b).

However, for $\Delta_{\rm c} = -\Delta/2$, the two quasienergies
$\epsilon_1^-$ and $-\epsilon_2^-$ degenerate. At this point, which marks the
center of the crossing, the number of different frequencies in the three-level
dynamics reduces to two again. This restores the familiar coherent tunneling
in the sense that there is again a simple periodic exchange of probability
between the regular regions\cite{lat}. However, the rate is much larger if
compared to the situation far off the crossing, and the chaotic region is now
temporarily populated during each probability transfer, twice per tunneling
cycle (Fig.~\ref{fig:tun3s}c).

In order to illustrate the above three-state model and to demonstrate its
adequacy, we have numerically studied a singlet-doublet crossing that occurs
for the double-well potential, Eq.~(\ref{eq:doublewell}), with $D/\hbar = 4$,
at a driving frequency $\omega = 0.982$ and amplitude $F = S/\sqrt{8D} =
0.015029$ (Figs.~\ref{fig:num3s}, \ref{fig:splitting}).
A comparison (not shown) of the appropriately scaled three-state theory 
(Fig.~\ref{fig:theo3s}) with this real singlet-doublet crossing
(Fig.~\ref{fig:num3s}) shows satisfactory agreement. Note that in the
real crossing, the quasienergy of the chaotic singlet {\it decreases\/} as
a function of $F$, so that the exact crossing occurs to the {\it left\/} of
the avoided one.

\section{Incoherent quantum dynamics}
\label{sec:diss}
\subsection{Master equation}
\subsubsection{System-bath model}
To achieve a microscopic model of dissipation, we couple the system
(\ref{eq:doublewell}) bilinearly to a bath of non-interacting harmonic
oscillators\cite{zwa}. The total Hamiltonian of system and bath is
then given by
\begin{equation}
H(t) = H_{\rm DW}(t)+
\sum_{\nu=1}^\infty \left(\frac{p_\nu^2}{2m_\nu}+\frac{m_\nu}{2}\omega_\nu^2
\left(x_\nu - \frac{g_\nu}{m_\nu\omega_\nu^2}x \right)^2 \right).
\end{equation}
We here couple the position $x$ of the system to an ensemble of
oscillators with masses $m_\nu$, frequencies $\omega_\nu$, momenta $p_\nu$,
and coordinates $x_\nu$, with the coupling strength $g_\nu$. The bath is fully
characterized by the spectral density of the coupling energy,
\begin{equation}
\label{SpectralDensity}
J(\omega) = \pi\sum_{\nu=1}^\infty \frac{g_\nu^2}{2m_\nu\omega_\nu}
\delta(\omega-\omega_\nu).
\end{equation}
For the time evolution we choose an initial condition of the Feynman-Vernon
type: at $t=t_0$, the bath is in thermal equilibrium and uncorrelated
to the system, i.e.,
\begin{equation}
\label{FVinitial}
\rho(t_0) = \rho_{\rm S}(t_0)\otimes\rho_{\rm B,eq},
\end{equation}
where $\rho_{\rm B,eq}=\exp(-\beta H_{\rm B})/{\rm tr}_{\rm B}
\exp(-\beta H_{\rm B})$
is the canonical ensemble of the bath and $1/\beta = k_{\rm B}T$.

Due to the bilinearity of the system-bath coupling, one can always eliminate
the bath variables to get an exact, closed integro-differential equation for
the reduced density matrix $\rho_{\rm S}(t)={\rm tr}_{\rm B}\rho(t)$, which
describes the dynamics of the central system, subject to dissipation
\cite{haa}.

\subsubsection{Born-Markov approximation}
\label{diss:markov}
%
In most cases, however, the integro-differential equation for $\rho_{\rm
S}(t)$ can be solved only approximately. In particular, in the limit of weak
coupling,
\begin{eqnarray}
\gamma & \ll & k_{\rm B}T/\hbar,\\
\gamma & \ll & |\epsilon_\alpha-\epsilon_{\alpha'}|/\hbar,
\end{eqnarray}
it is possible to truncate the time-dependent perturbation expansion in the
system-bath interaction after the second-order term. The quantity $\gamma$, to
be defined below, denotes the effective damping of the dissipative system, and
$|\epsilon_\alpha-\epsilon_{\alpha'}|/\hbar$ are the transition frequencies of
the central system. In the present case, the central system is understood to
include the driving\cite{blu1,blu2,gra,koh}, so that the transition
frequencies are given by quasienergy differences. The autocorrelations of the
bath decay on a time scale $\hbar/k_{\rm B}T$ and thus in the present limit,
instantaneously on the time scale $1/\gamma$ of the system correlations.  With
the initial preparation (\ref{FVinitial}), the equation of motion for the
reduced density matrix in this approximation is given by\cite{koh}
\begin{eqnarray}
\nonumber
\dot \rho_{\rm S}(t)
&=& -\frac{{\rm i}}{\hbar}\left[ H_{\rm S}(t),\rho_{\rm S}(t) \right]
+\frac{1}{\pi\hbar}\int_{-\infty}^\infty {\rm d}\omega\, J(\omega)n_{\rm th}
(\hbar\omega) \nonumber\\
&&\times \int_0^\infty {\rm d}\tau\left( {\rm e}^{{\rm i}\omega\tau}
\left[ \tilde x(t-\tau,t)\rho_{\rm S}(t),x\right] + \text{H.c.}\right)
\label{MasterEquationGeneral}
\end{eqnarray}
where $\tilde x(t',t)$ denotes the position operator in the interaction 
picture defined by
\begin{equation}
\tilde x(t',t) = U^\dagger(t',t)\, x\, U(t',t),
\end{equation}
with $U(t',t)$, the propagator of the conservative driven double well, given
in Eq.~(\ref{eq:floquetprop}). `H.c.' denotes the Hermitian conjugate and
\begin{equation}
n_{\rm th}(\epsilon)
=\frac{1}{{\rm e}^{\epsilon/k_{\rm B}T} - 1}=-n_{\rm th}(-\epsilon) - 1
\end{equation}
is the thermal occupation of the bath oscillator with energy $\epsilon$.  To
achieve a more compact notation, we require $J(-\omega)=-J(\omega)$.  In the
following, we shall restrict ourselves to an Ohmic bath, $J(\omega) =
m\gamma\omega$. This defines the effective damping constant $\gamma$.

We use the time-periodic components $|\phi_\alpha(t)\rangle$ of the Floquet
states as a basis to expand the density operator,
Eq.~(\ref{MasterEquationGeneral}). Expressing the matrix elements
\begin{equation}
X_{\alpha\beta}(t) = \langle\phi_\alpha(t)|x|\phi_\beta(t)\rangle
\end{equation}
of the position operator by their Fourier coefficients
\begin{eqnarray}
X_{\alpha\beta,n}
&=& \langle\langle\phi_\alpha(t)|x\,{\rm e}^{-{\rm i}n\omega t}
|\phi_\beta(t)\rangle\rangle = X_{\beta\alpha,-n}^*\, , \\
X_{\alpha\beta}(t) &=& \sum_n {\rm e}^{{\rm i}n\omega t}X_{\alpha\beta,n}\, ,
\end{eqnarray}
yields for the matrix elements $\sigma_{\alpha\beta}$ of the reduced
density matrix $\rho_{\rm S}$ the equation of motion\cite{dit1,dit2,blu2,koh}
\begin{eqnarray}
\dot\sigma_{\alpha\beta}(t)
&=& {{\rm d}\over{\rm d}t}\langle\phi_\alpha(t)|\rho_{\rm S}(t)|
    \phi_\beta(t)\rangle \nonumber\\
&=& -\frac{\rm i}{\hbar}(\epsilon_\alpha-\epsilon_\beta)
    \sigma_{\alpha\beta}(t) \nonumber\\
&&  +\sum_{\alpha'\beta'nn'}\big( N_{\alpha\alpha',n}X_{\alpha\alpha',n}
    \sigma_{\alpha'\beta'}X_{\beta'\beta,n'}  \nonumber\\
&&  -N_{\alpha'\beta',n}X_{\alpha\alpha',n'}X_{\alpha\alpha',n}
    \sigma_{\beta'\beta}\big) {\rm e}^{{\rm i}(n+n')\omega t}
    +\text{H.c.}\, .
\label{MasterEquationFloquet}
\end{eqnarray}
Note that the coefficients of this differential equation are periodic
in time with the period of the driving.
The $N_{\alpha\beta,n}$ are given by
\begin{equation}
N_{\alpha\beta,n} = N(\epsilon_\alpha-\epsilon_\beta+n\hbar\omega),\quad
N(\epsilon) = \frac{m\gamma\epsilon}{\hbar^2} n_{\rm th}(\epsilon).
\end{equation}
For $\epsilon\gg k_{\rm B}T$, $N(\epsilon)$ approaches zero.

Since the position operator $x$ is odd under transformation with the
generalized parity~(\ref{parity}), the master equations
(\ref{MasterEquationGeneral}) and (\ref{MasterEquationFloquet}) are invariant
under transformation with ${\sf P}_\omega$. Therefore, both the conservative
and the dissipative dynamics preserve the parity of the operator
$|\phi_\alpha\rangle\langle\phi_\beta|$. It is even if $|\phi_\alpha\rangle$
and $|\phi_\beta\rangle$ belong to the same parity class and odd otherwise.
Note that in particular, the projectors
$|\phi_\alpha\rangle\langle\phi_\alpha|$ and thus all density matrices
diagonal in the Floquet basis are even under ${\sf P}_\omega$.

\subsubsection{Rotating-wave approximation}
%
Assuming that dissipative effects are relevant only on a time scale much
larger than the period $2\pi/\omega$ of the driving, we average the
coefficients of the master equation (\ref{MasterEquationFloquet}) over
$2\pi/\omega$ and end up with the equation of motion
\begin{equation}
\dot\sigma_{\alpha\beta}(t)
=-\frac{\rm i}{\hbar}(\epsilon_\alpha-\epsilon_\beta)\sigma_{\alpha\beta}(t)
+\sum_{\alpha'\beta'} L_{\alpha\beta\alpha'\beta'}\sigma_{\alpha'\beta'} ,
\end{equation}
with the time-independent dissipative part
\begin{eqnarray}
L_{\alpha\beta\alpha'\beta'}
&=& \sum_n \left( N_{\alpha\alpha',n} + N_{\beta\beta',n}\right)
    X_{\alpha\alpha',n} X_{\beta'\beta,-n}
\nonumber\\
&& -\delta_{\beta\beta'}\sum_{\beta'',n}
   N_{\beta''\alpha',n}X_{\alpha\beta'',-n}
   X_{\beta''\alpha',n}\nonumber\\
&& -\delta_{\alpha\alpha'}\sum_{\alpha''n}
   N_{\alpha''\beta',n}
   X_{\beta'\alpha'',-n}X_{\alpha''\beta,n} .
\label{MasterEquationRWA}
\end{eqnarray}
This step amounts effectively to a rotating-wave approximation. It is,
however, less restrictive than the rotating-wave approximation introduced in
\cite{blu1,blu2} where dissipative effects are averaged over the generally
longer time scale $\text{max}_{\alpha,\beta,n}(2\pi\hbar
/(\epsilon_\alpha-\epsilon_\beta+n\hbar\omega))$.

\subsection{Dissipative chaos-assisted tunneling}
\label{diss:cat}

The crucial effect of dissipation on a quantum system is the disruption of
coherence: a coherent superposition evolves into an incoherent mixture.  Thus,
phenomena based on coherence, such as tunneling, are rendered transients that
fade out on a finite time scale $t_{\rm decoh}$. In general, for driven
tunneling in the weakly damped regime, this time scale gets shorter for higher
temperatures, reflecting the growth of transition rates\cite{hae2}. However,
there exist counterintuitive effects. For example, in the vicinity of an exact
crossing of the ground-state quasienergies, the coherent suppression of
tunneling\cite{gro1} can be stabilized with higher temperatures
\cite{dit1,dit2,mak} until levels outside the doublet start to play a r\^ole.
We have studied dissipative chaos-assisted tunneling numerically, using again
the real singlet-doublet crossing introduced in Sect.~\ref{sec:3s} (cf.\
Fig.~\ref{fig:num3s}) as our working example.

\begin{figure}
\centerline{
\psfig{width=8cm,figure=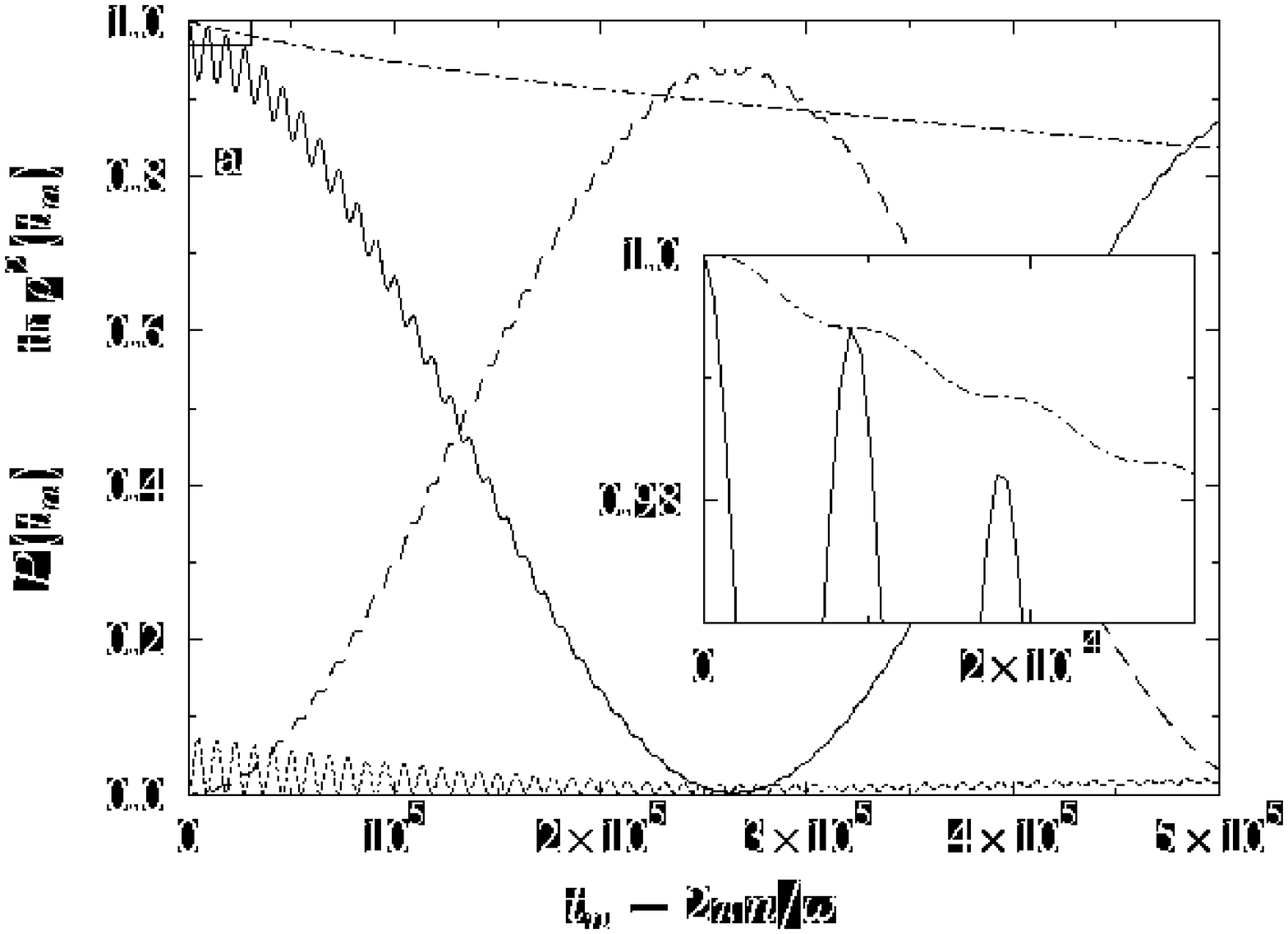}
\psfig{width=8cm,figure=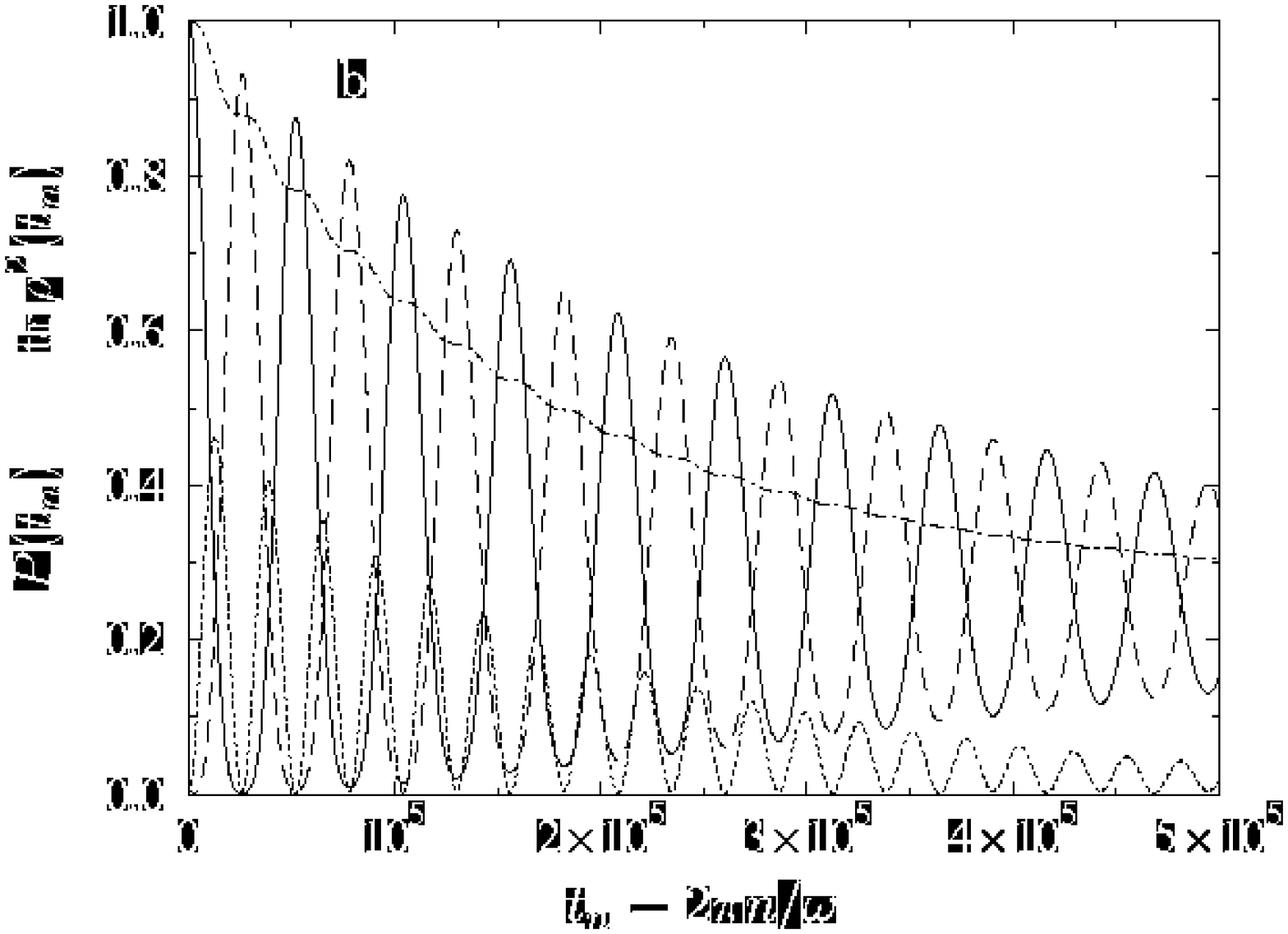}
}
\vspace{0.5cm}
\caption{
Occupation probabilities as in Fig.~\protect\ref{fig:tun3s}a,c, but in the
presence of dissipation. The dash-dotted line shows the time evolution of
${\rm tr}\,\rho^2$. The parameter values are $D/\hbar=4$, $\omega=0.982$,
$\gamma=10^{-6}$, $k_{\rm B}T/\hbar=10^{-4}$, and $F=0.0145$ (a),
$0.015029$ (b). The inset in (a) is a blow up of the rectangle in the upper
left corner of that panel.
}
\label{fig:diss:short}
\end{figure}

\begin{figure}
\centerline{
\psfig{width=8cm,figure=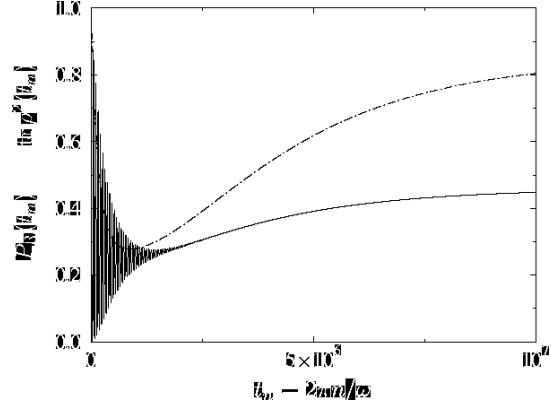}
}
\vspace{0.5cm}
\caption{
Time evolution of the return probability $P_{\rm R}$ (full line) and the 
coherence function ${\rm tr}\,\rho^2$ (dash-dotted) during loss and regain
of coherence. The parameter values are as in Fig.\
\protect\ref{fig:diss:short}b.
}
\label{fig:diss:recoherence}
\end{figure}

In the vicinity of a singlet-doublet crossing, the tunnel splitting {\it
increases\/} significantly---the essence of chaos-assisted tunneling. During
the tunneling, the chaotic singlet becomes populated periodically with
frequency $|\epsilon_2^--\epsilon_1^-|/\hbar$, cf.\ Eq.~(\ref{eq:tun3s}) and
Fig.~\ref{fig:tun3s}. The high mean energy of this singlet results in an
enhanced decay of coherence at times when $|\psi_{\rm c}\rangle$ is populated
(Fig.~\ref{fig:diss:short}). For the relaxation towards the asymptotic state,
also the slower transitions within doublets are relevant. Therefore, the
corresponding time scale $t_{\rm relax}$ can be much larger than 
$t_{\rm decoh}$ (Fig.~\ref{fig:diss:recoherence}).

To obtain quantitative estimates for the dissipative time scales, we
approximate $t_{\rm decoh}$ by the decay rate of ${\rm tr}\,\rho^2$, a measure
of coherence, averaged over time,
\begin{eqnarray}
\frac{1}{t_{\rm decoh}} 
&=& -\frac{1}{t}\int_0^t {\rm d}t' \frac{\rm d}{{\rm d}t'}\,{\rm tr}\,\rho^2(t')
    \\
&=& \left.\frac{1}{t}\left( {\rm tr}\,\rho^2(0)
- {\rm tr}\,\rho^2(t)\right)
\right. .
\end{eqnarray}
Because of the stepwise decay of the coherence (Fig.~\ref{fig:diss:short}),
we have chosen the propagation time $t$ as an $n$-fold multiple of the
duration $2\pi/|\epsilon_2^- - \epsilon_1^-|$ of the chaotic beats.  For this
procedure to be meaningful, $n$ should be so large that the coherence decays
substantially during the time $t$ (in our numerical studies to a value of
approximatly 0.9).
The time scale $t_{\rm relax}$ of the approach to the asymptotic state is
given by the reciprocal of the smallest real part of the eigenvalues of
the dissipative kernel.

\begin{figure}
\centerline{
\psfig{width=8cm,figure=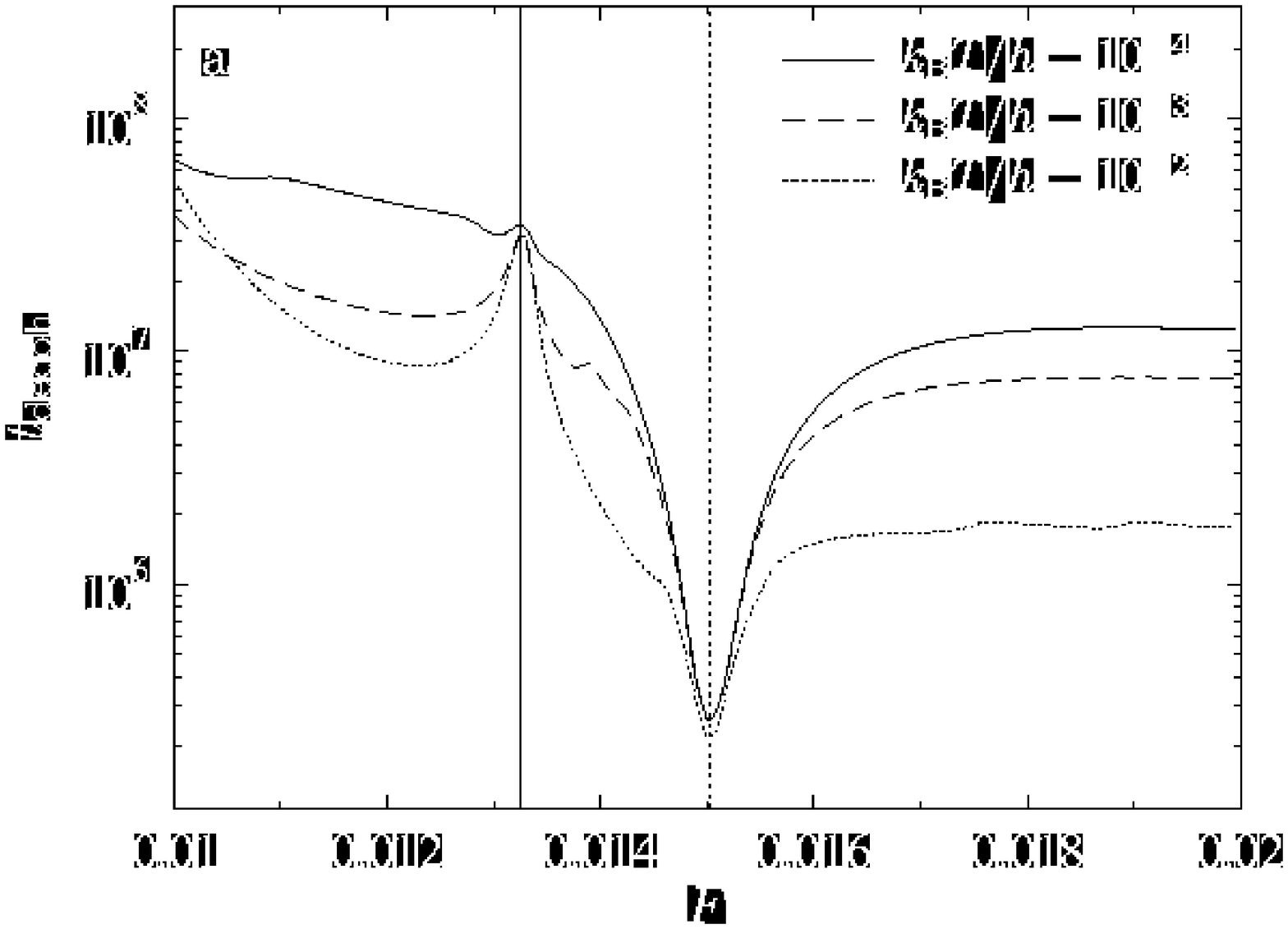}
\psfig{width=8cm,figure=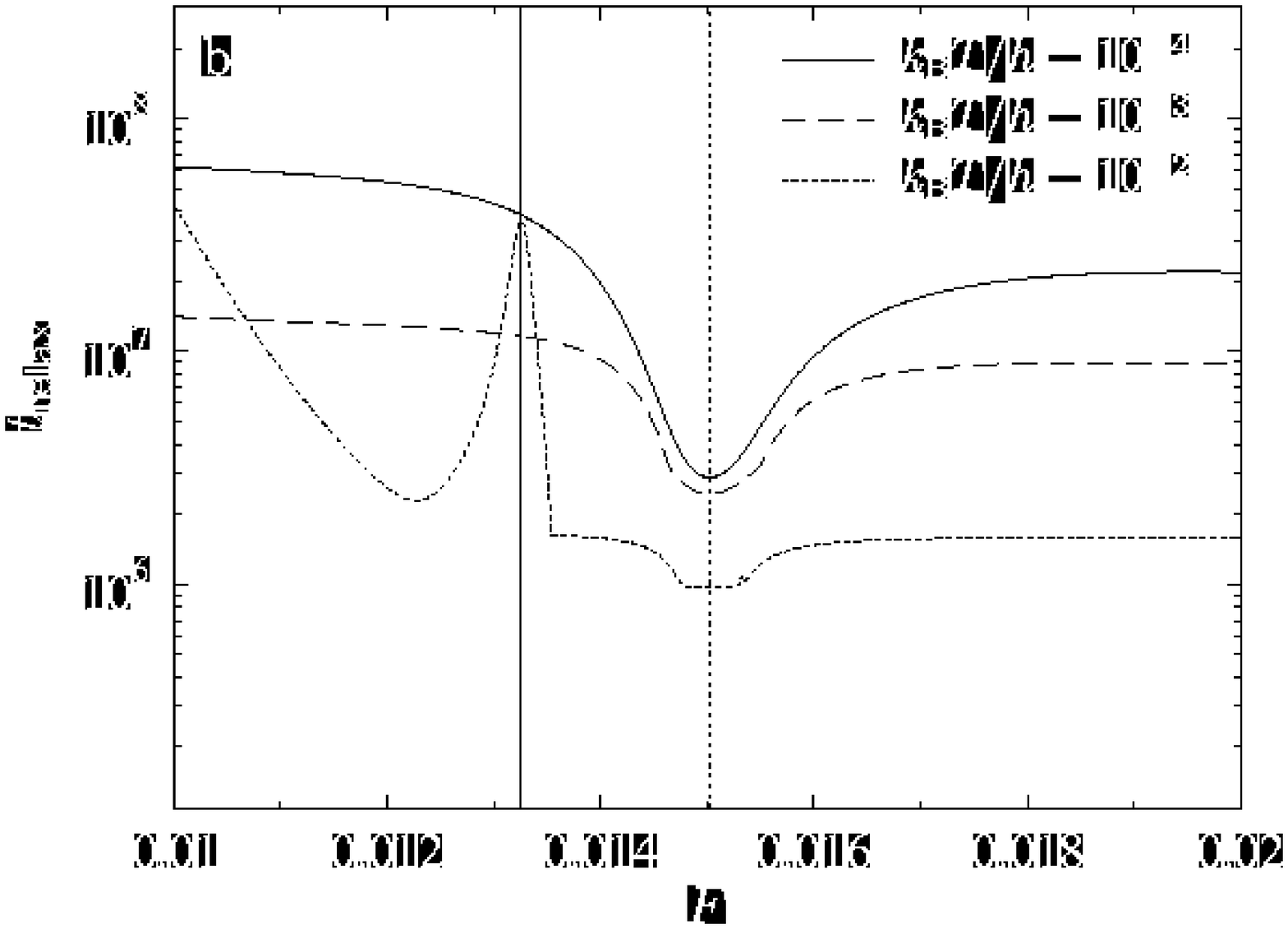}
}
\vspace{0.5cm}
\caption{
Time scales of the decay of ${\rm tr}\,\rho^2$ (a) and of the relaxation
towards the asymptotic solution (b) near the singlet-doublet crossing.  Near
the exact crossing ($F\approx 0.013$, full vertical line) coherence is
stabilized, whereas at the center of the avoided crossing ($F\approx 0.015$,
dashed vertical line) the decay of coherence is accelerated. The parameter
values are $D/\hbar=4$, $\omega=0.982$, $\gamma=10^{-6}$, temperature as given
in the legend.
}
\label{fig:diss:timescales}
\end{figure}

Outside the singlet-doublet crossing we find that the decay of coherence and
the relaxation take place on roughly the same time scale
(Fig.~\ref{fig:diss:timescales}). At $F \approx 0.013$, the chaotic singlet
induces an exact crossing of the ground-state quasienergies (see Fig.\
\ref{fig:splitting}), resulting in a stabilization of coherence with
increasing temperature.  At the center of the avoided crossing, the decay of
coherence becomes much faster and is essentially independent of
temperature. This indicates that transitions from states with mean energy far
above the ground state play a crucial r\^ole.

\subsection{Asymptotic state}
\label{diss:attractor}

As the dynamics described by the master equation (\ref{MasterEquationGeneral})
is dissipative, it converges in the long-time limit to an asymptotic state
$\rho_\infty(t)$. In general, this attractor remains time dependent but shares
all the symmetries of the central system, i.e.\ here, periodicity and
generalized parity. However, the coefficients of the master equation
(\ref{MasterEquationRWA}) for the matrix elements $\sigma_{\alpha\beta}$,
valid within a rotating-wave approximation, are time independent and so the
asymptotic solution also is. This means that we have eliminated the explicit
time dependence of the attractor by representing it in the Floquet basis and
introducing a mild rotating-wave approximation.

To gain some qualitative insight into the asymptotic solution, we focus on
the diagonal elements
\begin{equation}
L_{\alpha\alpha\alpha'\alpha'}
=2\sum_n N_{\alpha\alpha',n}|X_{\alpha\alpha',n}|^2,\quad \alpha\neq\alpha',
\label{fullRWA}
\end{equation}
of the dissipative kernel. They give the rates of the direct transitions from
$|\phi_{\alpha'}\rangle$ to $|\phi_\alpha\rangle$. Within a cruder
rotating-wave approximation\cite{blu2}, these are the only non-vanishing
contributions to the master equation which affect the diagonal elements
$\sigma_{\alpha\alpha}$ of the density matrix.

In the case of zero driving amplitude, the Floquet states
$|\phi_\alpha\rangle$ reduce to the eigenstates of the undriven Hamiltonian
$H_{\rm DW}$. The only non-vanishing Fourier component is then
$|c_{\alpha,0}\rangle$, and the quasienergies $\epsilon_\alpha$ reduce to the
corresponding eigenenergies $E_\alpha$. Thus $L_{\alpha\alpha\alpha'\alpha'}$
consists of a single term proportional to $N(\epsilon_\alpha -
\epsilon_{\alpha'})$ only. It describes two kinds of thermal transitions:
decay to states with lower energy and, if the energy difference is less than
$k_{\rm B}T$, thermal activation to states with higher energy. The ratio of
the direct transitions forth and back then reads
\begin{equation}
\frac{L_{\alpha\alpha\alpha'\alpha'}}{L_{\alpha'\alpha'\alpha\alpha}}
=\exp\left(-{(\epsilon_\alpha-\epsilon_{\alpha'}) \over k_{\rm B}T}\right).
\end{equation}
We have detailed balance. Therefore, the steady-state solution
$\sigma_{\alpha\alpha'}(\infty) \sim \exp(-\epsilon_\alpha/k_{\rm B}T)\,
\delta_{\alpha\alpha'}$. In particular, the occupation probability decays
monotonically with the energy of the eigenstates. In the limit $k_{\rm B}T\to0$,
the system tends to occupy the ground state only.

For a strong driving, each Floquet state $|\phi_\alpha\rangle$ contains a
large number of Fourier components and $L_{\alpha\alpha\alpha'\alpha'}$ is
given by a sum over contributions with quasienergies $\epsilon_\alpha -
\epsilon_{\alpha'} + n\hbar\omega$. Thus a decay to states with ``higher''
quasienergy (recall that quasienergies do not allow for a global
ordering) becomes possible due to terms with $n<0$. Physically, they describe
dissipative transitions under absorption of driving-field quanta.
Correspondingly, the system tends to occupy Floquet states comprising many
Fourier components with low index $n$. According to
Eq.~(\ref{eq:meanen:fourier}), these states have low mean energy.

\begin{figure}
\centerline{
\psfig{width=8cm,figure=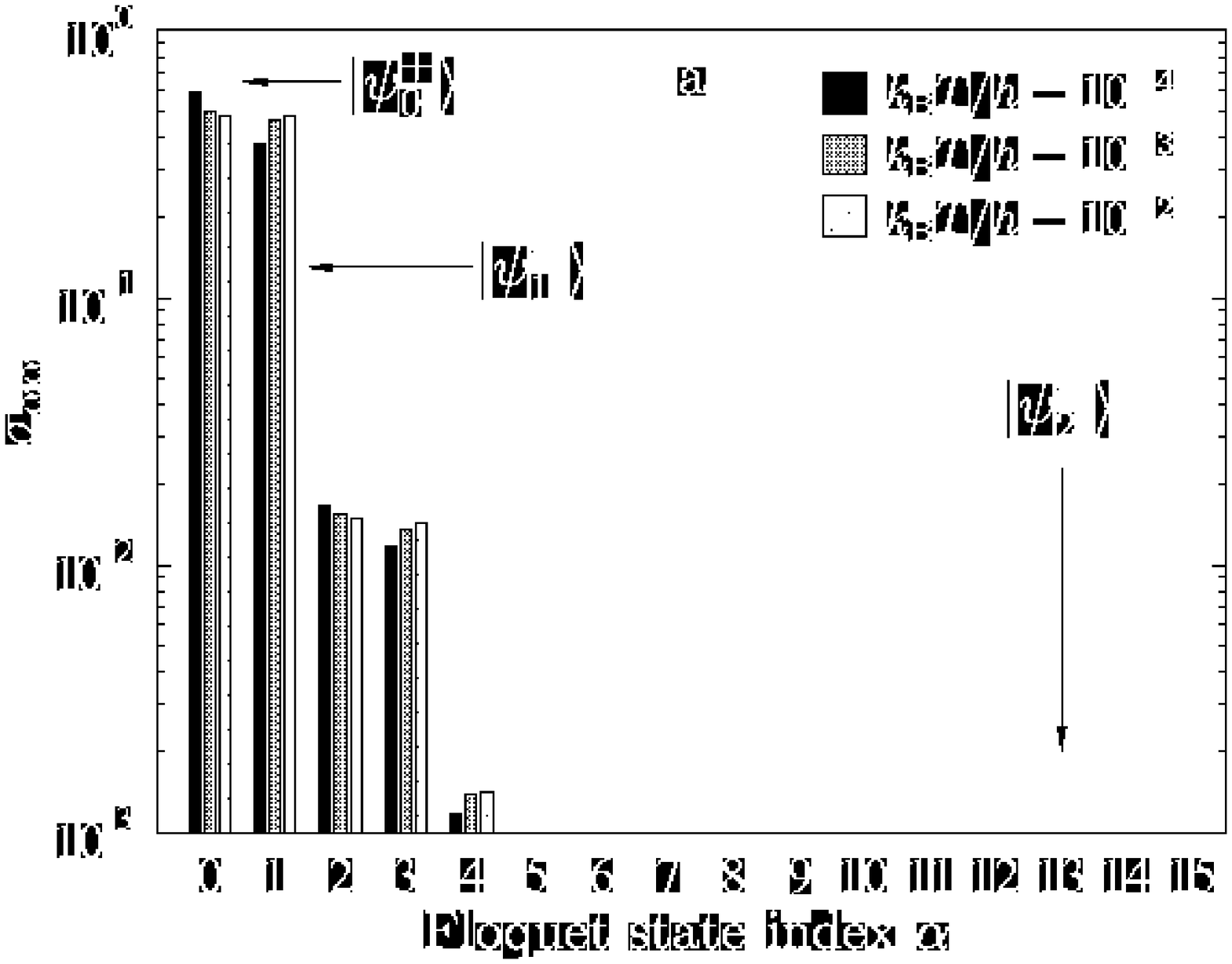}
\psfig{width=8cm,figure=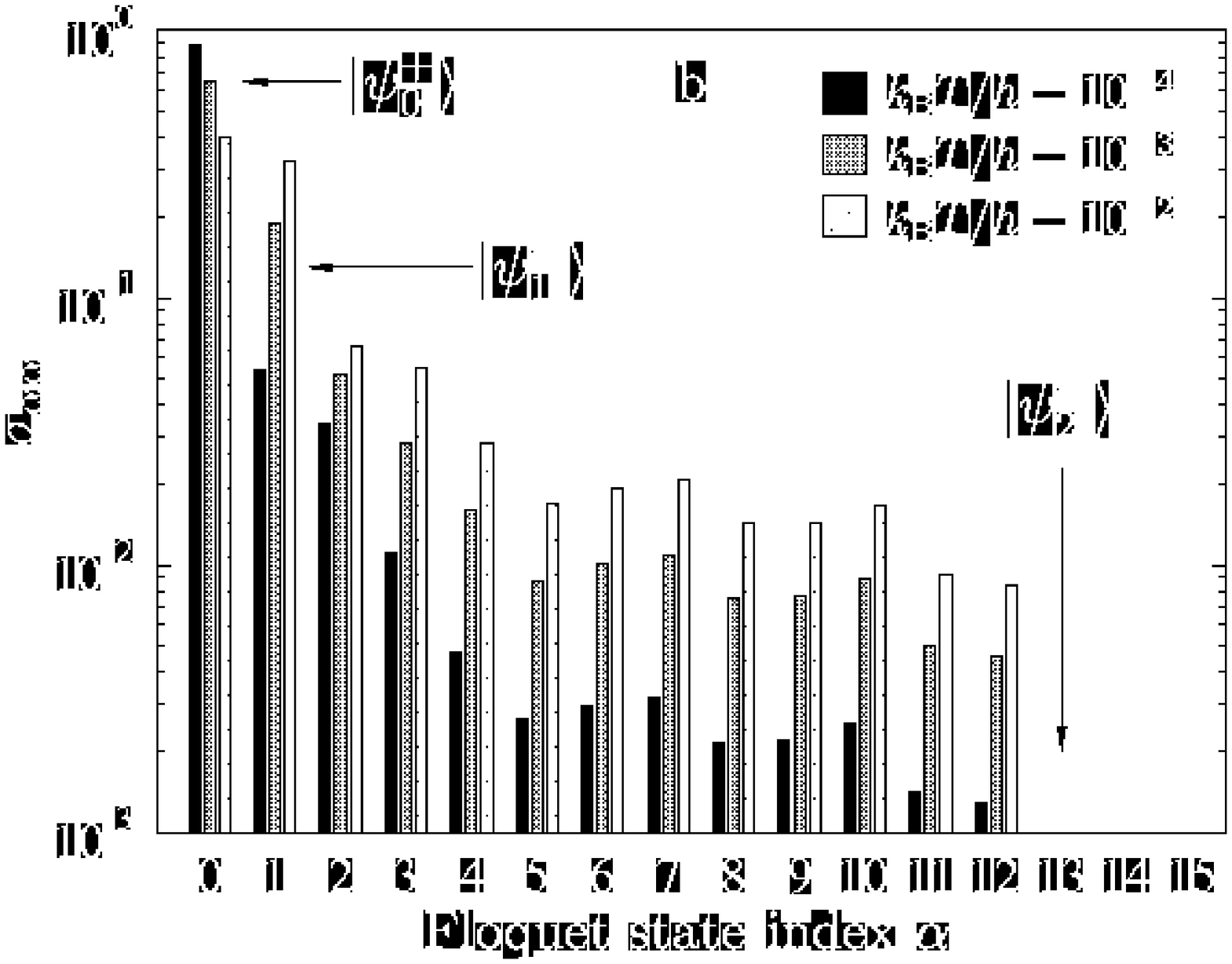}
}
\centerline{
\psfig{width=8cm,figure=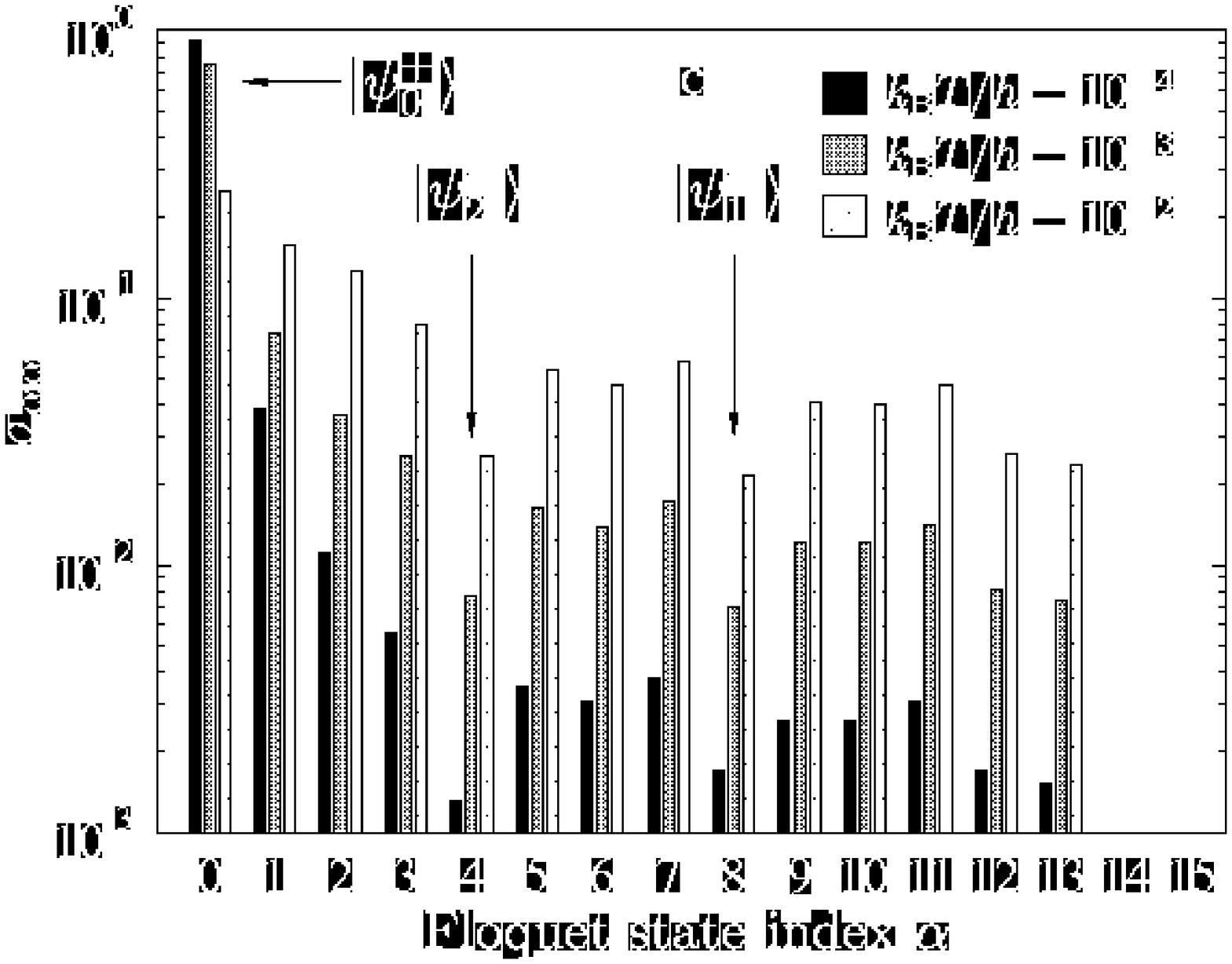}
}
\vspace{0.5cm}
\caption{
Occupation probability $\sigma_{\alpha\alpha}$ of the Floquet states
$|\psi_\alpha\rangle$ in the long-time limit.
The parameter values are $D/\hbar = 4$, $\omega = 0.982$, $\gamma=10^{-6}$,
and $F = 0.013$ (a), $0.0145$ (b), $0.015029$ (c), temperature as given in
the legend.
}
\label{fig:occupation}
\end{figure}

The effects under study are found for a driving with a frequency of the order
of unity. Thus for a quasienergy doublet, i.e., far off the three-level
crossing, we have $|\epsilon_\alpha-\epsilon_{\alpha'}| \ll \hbar\omega$, and
$L_{\alpha'\alpha'\alpha\alpha}$ is dominated by contributions with $n<0$,
where the splitting has only small influence. However, as a consequence of
symmetry, the splitting is the main difference between the two partners of the
quasienergy doublet. Therefore, with respect to dissipation, both should
behave similarly. In particular, one expects an equal population of the
doublets even in the limit of zero temperature (Fig.~\ref{fig:occupation}a).
This is in contrast to the undriven case.

In the vicinity of a singlet-doublet crossing the situation is more subtle.
Here, the odd partner, say, of the doublet mixes with a chaotic singlet, cf.\
Eq.\ (\ref{eq:psi012}), and thus acquires components with higher energy.  Due
to the high mean energy $E_{\rm c}^-$ of the chaotic singlet, close to the top
of the barrier, the decay back to the ground state can also proceed indirectly
via other states with mean energy below $E_{\rm c}^-$.  Thus
$|\phi_1^-\rangle$ and $|\phi_2^-\rangle$ are depleted and mainly
$|\phi_0^+\rangle$ will be populated.  However, if the temperature is
significantly above the splitting of the avoided crossing, thermal activation
from $|\phi_0^+\rangle$ to $|\phi_{1,2}^-\rangle$, accompanied by depletion
via the states below $E_{\rm c}^-$, becomes possible.  Thus asymptotically,
all these states become populated in a steady flow
(Figs.~\ref{fig:occupation}b,c).

\begin{figure}
\centerline{
\psfig{width=8cm,figure=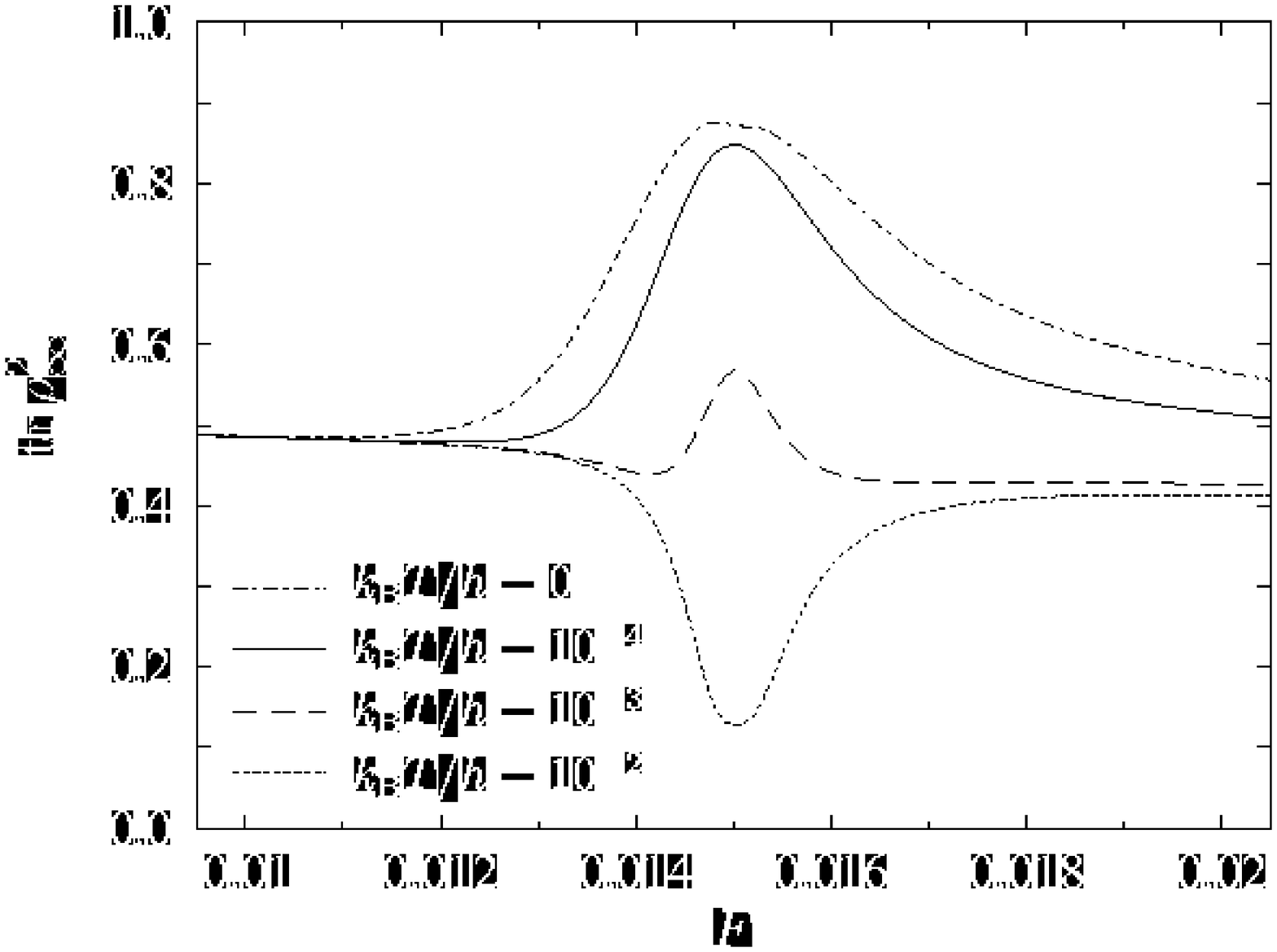}
}
\vspace{0.5cm}
\caption{
Coherence of the asymptotic state in the vicinity of a singlet-doublet
crossing for different temperatures as given in the legend. The other
parameter values are $D/\hbar = 4$, $\omega = 0.982$, and $\gamma=10^{-6}$.
}
\label{fig:cic}
\end{figure}

\begin{figure}
\centerline{
\psfig{width=8cm,figure=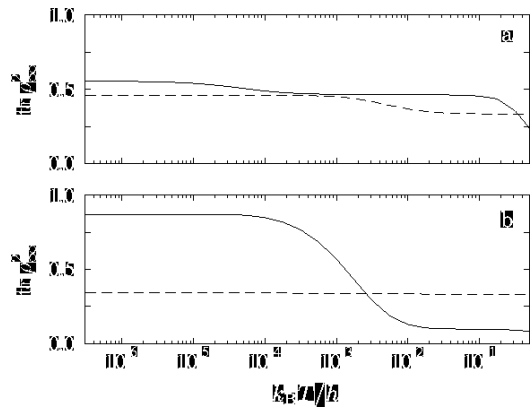}
}
\vspace{0.5cm}
\caption{
Coherence of the asymptotic state in the vicinity of a singlet-doublet
crossing for $F=0.013$ (a) and $F=0.015029$ (b): exact calculation (full line)
compared to the values resulting from a three-level description (dashed) of
the dissipative dynamics. The other parameter values are $D/\hbar=4$,
$\omega = 0.982$, and $\gamma=10^{-6}$.
}
\label{fig:coh}
\end{figure}

An important global characteristic of the asymptotic state is its coherence
${\rm tr}\,\rho_\infty^2$. Its value gives approximately the reciprocal of the
number of incoherently occupied states. It equals unity only if the attractor
is a pure state. According to the above scenario, we expect ${\rm
tr}\,\rho_\infty^2$ to assume the value $1/2$, in a regime with strong driving
but preserved doublet structure, reflecting the incoherent population of the
ground-state doublet.  In the vicinity of the singlet-doublet crossing where
the doublet structure is dissolved, its value should be close to unity for
temperatures $k_{\rm B}T\ll b$ and much less than unity for $k_{\rm B}T\gg b$
(Figs.~\ref{fig:cic}, \ref{fig:coh}). This means that the crossing of the
chaotic singlet with the regular doublet leads to an improvement of coherence
if the temperature is below the splitting of the avoided crossing, and a loss
of coherence for temperatures above the splitting. This phenomenon amounts to
a {\it chaos-induced coherence\/} or incoherence, respectively.

The crucial r\^ole of the decay via states not involved in the three-level
crossing can be demonstrated by comparing it with the dissipative dynamics
including only these three levels (plus the bath). At the crossing, the
three-state model results in a completely different type of asymptotic state
(Fig.~\ref{fig:coh}). The failure of the three-state model in the presence of
dissipation clearly indicates that in the vicinity of the singlet-doublet
crossing, it is important to take a large set of levels into account.

\section{Conclusion}
\label{sec:conc}
Nonlinear systems with a mixed phase space still represent a formidable
challenge for a theoretical understanding. On the classical level, the
intricate fractal interweaving of regular and chaotic regions of phase space
is far from being exhaustively studied. Quantum-mechanical uncertainty reduces
the richness of the classical phase-space structure, but at the same time,
coherence effects like tunneling add new elements to the dynamics. For
example, in a bistable system, a driving with an energy in the regime between
the unperturbed tunnel splittings and the typical separation of doublets leads
to transport phenomena that combine tunneling and classical chaotic diffusion
as essential elements. Yet another basic energy scale enters if the
unavoidable coupling of the system to its environment is taken into account.
If the finite width the levels attain in the presence of dissipation is
comparable, in turn, to the energies characterizing driven tunneling,
unfamiliar phenomena due to the interplay of chaos, tunneling, and decoherence
occur in the transient dynamics and possibly also in the asymptotic state of
the driven damped quantum system.

In this paper, we have selected a specific case out of the host of problems to
be studied in this field: dissipative chaotic tunneling in the vicinity of
crossings of chaotic singlets with tunnel doublets. In order to obtain a firm
quantitative basis, we have endowed a prototypical model for chaotic
tunneling, a harmonically driven double well, with dissipation.  Thereby, we
have followed the usual approach of coupling to a heat bath, but adapted to
the periodic time dependence of the central system. Moreover, we have
carefully avoided to destroy, by the approximations introduced in the
derivation of a master equation, the specific spectral characteristics of
chaotic tunneling that we are interested in. As a simple intuitive model to
compare against, we have constructed a three-state system which in the case of
vanishing dissipation, provides a faithful description of an isolated
singlet-doublet crossing.

As an example for the interplay of chaotic tunneling with dissipation, we
mention an effect that might be termed ``chaos-induced coherence'' for
short. Suppose that damping and temperature are such that far off the
singlet-doublet crossing, the level width is of the same order or larger than
the tunnel splitting. There, coherent tunneling is then largely suppressed
even as a transient. Close to the crossing, however, the widening of the
doublet enforced by the intersecting third level creates a separation of the
time scales of the coherent tunneling and of its decay due to incoherent
processes, without altering the dissipation as such. As a consequence, the
transient dynamics now exhibits many tunneling cycles, each of which includes
two ``stop overs'' in the chaotic phase-space region, one on the way to, one
on the way back.

The study of the asymptotic state, the ``quantum attractor'', demonstrates
clearly that a three-state model of the singlet-doublet crossing is no longer
adequate once dissipation is effective. This is so because the coupling to the
heat bath enables processes of decay and thermal activation that connect the
states in the crossing with other, ``external'' states of the central
system. In the presence of driving, the asymptotic state is no longer
literally a state of equilibrium. Rather, incoherent processes create a steady
flow of probability involving states within as well as outside the
crossing. As a result, the composition of the asymptotic state, expressed for
example by its coherence ${\rm tr}\,\rho_\infty^2$, will be markedly
different at the center of the crossing as compared to the asymptotic state
far away from the crossing, even if that is barely visible in the corresponding
phase-space structure.

Many more phenomena at the overlap of chaos, tunneling, and dissipation await
being unraveled. They include four-state crossings formed when two doublets
intersect, chaotic Bloch tunneling along extended potentials with a large
number of unit cells instead of just two, the influence of decoherence on a
multi-step mechanism of chaotic tunneling \cite{dor,fri} and transient
tunneling between coexisting strange attractors, to mention only a few. At the
same time, analytical tools more specifically taylored to the investigation of
the quantum-classical correspondence in mixed systems, such as a phase-space
entropy \cite{mir}, will provide additional insight.
\section*{Acknowledgments}
Financial support of this work by the Deutsche Forschungsgemeinschaft (grant
No.\ Di~511/2-2) is gratefully acknowledged. One of us (TD) would like to
thank for the warm hospitality enjoyed during a stay in the group of
Prof.~O.~Bohigas, Institut de Physique Nucl\'eaire, Orsay, financed by
Deutsche Forschungsgemeinschaft (grant No.\ Di~511/4-1), where this work was
completed.

%
%

\end{document}